\documentclass[fleqn,usenatbib]{mnras}

\usepackage{newtxtext,newtxmath}

\usepackage[T1]{fontenc}

\DeclareRobustCommand{\VAN}[3]{#2}
\let\VANthebibliography\thebibliography
\def\thebibliography{\DeclareRobustCommand{\VAN}[3]{##3}\VANthebibliography}

\usepackage{graphicx}	
\usepackage{amsmath}	
\usepackage{caption}
\usepackage{subcaption} 
\usepackage[normalem]{ulem}

\newcommand{\gcm}{g\,cm$^{-3}$}	
\newcommand{\kmskpc}{km\,s$^{-1}$\,kpc$^{-1}$} 
\newcommand{\kms}{km\,s$^{-1}$} 

\title[WD binary asteroid scattering]{Binary asteroid scattering around white dwarfs}

\author[McDonald \& Veras]{
Catriona H. McDonald,$^{1, 2}$\thanks{E-mail: catriona.mcdonald@warwick.ac.uk}
Dimitri Veras,$^{1, 2, 3}$
\\
$^{1}$Centre for Exoplanets and Habitability, University of Warwick, Coventry CV4 7AL, UK\\
$^{2}$Department of Physics, University of Warwick, Coventry CV4 7AL, UK \\
$^{3}$Centre for Space Domain Awareness, University of Warwick, Coventry CV4 7AL, UK \\
}

\date{Accepted XXX. Received YYY; in original form ZZZ}

\pubyear{2022}

\begin{document}
\label{firstpage}
\pagerange{\pageref{firstpage}--\pageref{lastpage}}
\maketitle

\begin{abstract}
Increasing observations of white dwarf atmospheric pollution and disrupting planetesimals is driving increased studies into the fate of exo-asteroids around post-main-sequence stars. 
Planetesimal populations in the Solar System which are most likely to survive the violent post-main-sequence evolution, such as the Kuiper Belt, display a large binary fraction with a propensity for near equal-mass components and provide a previously unexplored population of planetesimals which are likely to exist around white dwarfs. 
Here we simulate the dynamical evolution of equal-mass binary asteroid systems around white dwarfs using the N-body integrator \texttt{REBOUND} for $1$~Gyr. 
We confirm that giant planets are efficient at dissociating and ejecting binary asteroid systems on eccentric orbits, while Earth-mass planets are better at keeping planetesimals in their planetary systems. 
We find binary systems can be dissociated and ejected from their systems across Myr timescales, producing interstellar objects. 
We do not expect a population of free-floating binary asteroid systems as all ejected planetesimals are gravitationally unbound from each other. 
Further, we discuss the influence of asteroid binarity on the white dwarf pollution process and find there is little to no impact on how close a body can get to a star.
However, the orbital evolution of binary asteroids changes the distribution of planetesimals available in a white dwarf planetary system to be further scattered onto white dwarf polluting orbits.
\end{abstract}




\section{Introduction} \label{sec:intro}
White dwarfs provide a unique opportunity to understand the composition of planets and planetesimals in exoplanetary systems, as white dwarfs have such intense surface gravities that elements heavier than hydrogen and helium should quickly sink to the cores and hence no longer be visible in spectroscopic observations \citep{Paquette1986, Wyatt2014}.
However, between a quarter and half of all white dwarfs are observed with metals in their atmospheres \citep{Zuckerman2010, Koester2014, Kepler2015, Kepler2016, Coutu2019} which could only be present if they had recently be deposited on the surface.

A large proportion of this accreted material appears to be aligned with the composition of terrestrial planetesimals \citep{Zuckerman2010, Jura2014, Hollands2018, Doyle2019} and a much smaller proportion with more volatile rich bodies \citep{Farihi2013, Raddi2015, GentileFusillo2017, Xu2017, Hoskin2020}.
It is believed that this material originates in asteroidal bodies which have been perturbed onto orbits which cross the Roche limit of the white dwarf and hence are tidally disrupted. 
The subsequent material likely forms a disc of dust and/or gas orbiting around the white dwarf. 
Such dusty debris discs are observed in the form of infrared excesses around $1-3$ per cent of white dwarfs \citep{RebassaMansergas2019} and gaseous discs are observed around $0.04-0.1$ per cent of white dwarfs \citep{Manser2020}.
But it is expected that most white dwarfs which exhibit metallic pollution should also be host to debris discs (see section 4 and figure 9 of \cite{Bonsor2017}) with up to $90$ per cent of those discs being currently unobservable \citep{Rocchetto2015}.

As a star leaves the main sequence, its increased luminosity and physical size will have a considerable effect on the planetary system that might reside around it.
Close in planets will be engulfed by the expanding stellar envelope, and planets which escape this fate will be pushed out onto orbits $2-3\times$ larger than their main sequence distances \citep{Veras2016a}.
A significant proportion of minor bodies within $\sim 7$~au of the host star will be broken down to their strongest components through the increased YORP effect during the giant branch phases of stellar evolution \citep{Veras2014b}.
At larger distances, \cite{Veras2020e} find that asteroids with $a \gtrsim 50$~au will almost always avoid rotational fission due to the YORP effect. 
Further, planetesimals at these distances, such as in an exo-Kuiper belt, will likely remain intact through the post-main-sequence increased luminosity and dynamical instabilities \citep{Debes2002, Veras2015a, Veras2016b, Mustill2018, Maldonado2020a, Maldonado2020b, Maldonado2021}.

The increasing number of observations showing planetesimals in various stages of disruption transiting white dwarfs \citep{Vanderburg2015, Vanderbosch2020, Vanderbosch2021, Guidry2021, Farihi2022}, are presenting new challenges to our understanding of remnant planetary systems around white dwarfs. 
Thus using our knowledge of Solar System asteroids, which remain the most well studied population of planetesimals, could provide the key to improving our efforts to model white dwarf debris systems.

In the Solar System a significant proportion of planetesimals exist in binary, or higher multiplicity, systems. 
This proportion has been shown observationally to vary from several per cent to several tens of per cent depending on both location and planetesimal size, and theorised to be as high as 100 per cent at formation \citep{Fraser2017}.

In particular, photometric studies of the Main belt suggest that $6 \pm 3$ per cent of asteroids with radius larger than $10$~km should be binaries with comparable sized components \citep{Behrend2006}. 
The Near-Earth asteroid (NEA) population appears to host $15-17$ per cent of bodies larger than $0.3$~km in diameter as binaries \citep{Margot2002, Pravec2006} and a further $\sim 10$ per cent of NEA bodies exist as contact binaries \citep{Benner2006}. 
The prevalence of multiple asteroid systems persists throughout the outer regions of the Solar System with $6-10$ per cent of Trojan asteroids existing as contact binaries \citep{Mann2007} and three Trojans containing their own satellites \citep{Merline2002, Marchis2006, Noll2020b}.


Out to even further orbital distances, the Trans-Neptunian Objects (TNOs) in the Kuiper belt are estimated to have a $10-20$ per cent binary fraction \citep{Stephens2006, Noll2008}.
The cold classical Kuiper belt objects (CCKBOs) are a population of planetesimals harbouring low inclinations and eccentricities alongside red colours and high albedos \citep{Nesvorny2019} and are thought to be a primordial reservoir of planetesimals which formed at their current location. 
The binary fraction for the CCKBOs is thought to be higher than the TNO region as a whole at $\sim 20-30$ per cent \citep{Benecchi2019}.

It has been suggested that binarity is a natural consequence of planetesimal formation and thus populations such as the CCKBOs could have had a near $100$ per cent binary fraction at formation \citep{Fraser2017}.
Thus these bodies perhaps provide a unique opportunity to look at the relatively unprocessed results of planetesimal formation, without the influence of the Solar System's particular dynamical history. 

The planetesimals which are most likely to survive post-main-sequence evolution are those in the outer regions of planetary systems where binarity is high in the Solar System.
However, this potentially significant population has been almost completely ignored in previous post-main-sequence investigations.
Thus, in this paper we investigate the dynamical evolution of binary asteroids in white dwarf planetary systems, and characterise their influence on observable pollution, transiting debris and the population of interstellar free-floaters (of which 1I/2017 U1 'Oumuamua is an example).

In Section~\ref{sec:sims} we outline the set up of our simulations and the conditions for asteroid tidal disruption, system ejection and binary dissociation. 
We apply these simulations to a Solar System analogue including the four giant planets, as well as systems where the binary asteroids are interior and exterior to an Earth-mass planet's orbit, in Section~\ref{sec:results}.
In Section~\ref{sec:discussion} we discuss some of the implications of asteroid binarity on the prospects for white dwarf debris systems and the production of interstellar objects.
Finally, we conclude in Section~\ref{sec:conc}.

\section{\texttt{REBOUND} simulations} \label{sec:sims}
To investigate the fate of binary asteroids in a post-main-sequence planetary system, we carry out N-body simulations using the \texttt{REBOUND} package \citep{ReinLiu2012}. 

We utilise the \texttt{WHFast} integrator module within \texttt{REBOUND} \citep{ReinTamayo2015} which is an implementation of the symplectic Wisdom-Holman integrator \citep{WisdomHolman1991}.
We chose to use \texttt{WHFast} instead of an adaptive timestep integrator such as \texttt{IAS15} \citep{ReinSpiegel2015} as in order to capture the dynamics of the binary orbit alongside the circumstellar orbit the integration timestep needed to be small and we found a significant speed advantage in using \texttt{WHFast} over \texttt{IAS15} without loss of information. 

Although the use of a Wisdom Holman integration scheme such as \texttt{WHFast} is most accurate for systems where motion is dominated by a central potential and other perturbations are small \citep{ReinLiu2012}, we carried out a number of tests to ensure that our numerical simulations are sufficient and provide valid results.

Firstly, we confirm that the relative error on energy and angular momentum of the majority of our simulations are of the order $\times 10^{-10}$, and never larger than $\times 10^{-7}$, ensuring a reasonable accuracy for our numerical simulations.

Secondly, we carried out a small number of test simulations with varying timestep values to check our results converge.
For a total of 12 asteroid systems in each planetary architecture considered in this work, we carried out additional simulations with identical initial conditions and timesteps either a factor of three larger or one third smaller than dt~$\sim 0.006$~yr.
We find that for the Solar System analogue simulations (Section~\ref{subsec:SSAnalog}) and those containing planetesimals exterior to an Earth-mass planet (Section~\ref{subsec:earthmass_ext}), only a single binary asteroid system has a different fate dependent on the simulation timestep used.
The simulations with planetesimals interior to an Earth mass planet (Section~\ref{subsec:earthmass_int}) had no divergent results with differing timesteps.
As there is only a small change in results dependent on the timestep used, and using a timestep a factor of three smaller increases the computation time by a factor of five, we find that the simulation procedure outlined above is sufficient for this general study.

Finally, we compared the performance of the \texttt{WHFast} integrator to the standard \texttt{Leapfrog} scheme also provided in \texttt{Rebound} \citep{ReinLiu2012}.
We again ran a small number of simulations for each planetary architecture using the \texttt{Leapfrog} integrator and examined the relative energy and angular momentum errors and the convergence of the results.

This integrator showed a similar outcome in terms of energy and angular momentum, with relative errors on the order of $\times 10^{-10}$.
However, this integrator also produced significantly more divergent results than with \texttt{WHFast}. 
While considering the Solar System analogue architecture, 5 out of 12 simulations ran with the \texttt{Leapfrog} integrator produced divergent results dependent on the timestep chosen, compared to the single divergent result using \texttt{WHFast}.

Thus, we consider our choice of \texttt{WHFast} to be robust and sufficient for the aims of this study.

The outputs of each of the simulations carried out in this study were stored using the SimulationArchive format available in \texttt{REBOUND} which stores the state of the simulation regularly allowing for exact reproducibility \citep{ReinTamayo2017}. 

Modelling binary asteroids orbiting a star is computationally demanding due to the difference in timescales for the two orbits. 
Thus we sought to find the combination of integration timestep and SimulationArchive output which would allow reasonable integration times. 
In order to preserve the information provided by the binary orbit, the integration timestep was set at $\sim 0.006$~yr, the same order of magnitude as the period of the tightest binaries we consider and three orders of magnitude smaller than the period of the widest. 
A snapshot was saved to the SimulationArchive every $10^5$~yrs.
As the binary systems are not expected to interact with each other significantly, we chose to simulate 100 binary asteroid systems across 25 simulations each containing four binaries.
This provided a speed advantage over simulating 100 systems individually.
With this configuration simulating the evolution of 100 binary asteroids in a planetary system took $\sim 5200$~hrs.

The planet and planetesimal architectures used in the simulations in this paper are now explained.

\subsection{Simulation set up} \label{subsec:setup}
All of the simulations in this study began with the same basic set up. 
A central white dwarf was initialized with $M_\text{WD} = 0.6$~M$_\odot$, which corresponds to the peak of the white dwarf mass distribution \citep{Althaus2010, Kleinman2013, Tremblay2016, McCleery2020}.

We then added a number of binary asteroid systems orbiting around the central star. 
Each binary system is comprised of two equal size components with a mass of $m_\text{a} = 2.5 \times 10^{19}$~kg, calculated assuming that each component has a radius of $r_\text{a} = 125$~km and a fiducial density of $3$~\gcm{}. 
These values were motivated by the observation that $100$~km class CCKBOs have a large percentage of resolved binaries with near equal-sized components \citep{Parker2010, delaFuenteMarcos2017, Nesvorny2019} and that TNOs in the mass range $10^{17}-10^{22}$~kg are observed with calculated densities up to $4$~\gcm{} \citep{Carry2012}. 

A number of circumstellar semi-major axis ranges were considered based on the particular planetary system architectures which are described in Section~\ref{sec:results}.
The circumstellar orbit was further initiated with a random eccentricity between $0-1$ (unless otherwise specified) and random inclination between $0-1^\circ$. The initial mean anomaly of the circumstellar orbit was randomly chosen between $0-2\pi$ radians.
The initial longitudes of ascending node and pericentre are set as zero.
In order to implement our binary asteroid systems in \texttt{REBOUND} and correctly determine orbits, we set the primary of the system within the code as one of the binary asteroid components.
The other binary component is then designated as the secondary.

A large range of initial eccentricity values are chosen to cover the possibility that dynamical interactions with remnant planets may have already excited the circumstellar planetesimal orbit to higher eccentricities through a number of mechanisms \citep[e.g.][]{Wyatt2017, Pichierri2017}, including eccentricity excitation without destructive instabilities \citep{OConnor2022}.
 
The range of possible binary semi-major axes values were again motivated by observations.
\cite{Nesvorny2019} define the separation of a binary asteroid system as the ratio of the binary semi-major axis to the combined size of the binary components $a_\text{B}/R_\text{B}$, where $R_\text{B}^3 = R_1^{3} + R_2^{3}$, such that $R_1$, $R_2$ are the radii of the primary and secondary, respectively, and $a_{\text{B}}$ is the semi-major axis of the primary-secondary orbit.
Fig.~2 of \cite{Nesvorny2019} shows the distribution of binary separations for different dynamical classes of KBO using the catalogue of physical and orbital properties of binary asteroids maintained by W. R. Johnston on the NASA Planetary Data System (PDS) node \citep{Johnston2019}. 
From this figure, it can be identified that binary CCKBOs have a binary separation in the range $10 < a_\text{B}/R_\text{B} < 1000$.
Using the previously described asteroid properties chosen, we can thus find an approximate range of $a_\text{B}$ for our binary asteroid systems; $1500 < a_\text{B} < 1.5 \times 10^5$~km. 

Finally we add planets to the simulations based on the planetary architecture of interest as discussed in Section~\ref{sec:results}.
We now discuss the different outcomes for the binary asteroids in our simulations.

\subsection{Encounters with the white dwarf} \label{subsec:sim_min_exit}
We remove a body if it approaches within the white dwarf's Roche limit, at which point it would undergo tidal disruption. 

Here, we adopt the Roche radius expression assuming a solid, spinning rubble pile as in \cite{Veras2017},
\begin{equation}
    \label{eq:roche_rad}
    r_{\text{Roche}} = 0.89 \left( \frac{M_\text{WD}}{\rho_\text{a}} \right)^{1/3},
\end{equation}
where $M_\text{WD}$ is the mass of the white dwarf and $\rho_\text{a}$ is the bulk density of the orbiting body. 
We adopt a body density of $3$\gcm{} to represent a body largely consisting of solid ices and dust and use a fiducial white dwarf mass of $M_\text{WD} = 0.6M_\odot$, which results in $r_\text{Roche} = 0.94$~R$_\odot \sim 0.004$~au.

Due to the computational timestep limitations of these simulations and the high velocity of bodies on orbits with small pericentres, we also take into account the possibility that the bodies in our simulations could cross the Roche radius between timesteps.

For each recorded simulation snapshot we calculate the expected pericentre distance for each asteroid using the osculating orbital elements and identify if this is within $5$ per cent of the white dwarf's Roche radius.
When this occurs, we then return to the previous simulation archive step ($\sim 6 \times 10^{5}$~yrs before) and proceed to carry out a further simulation with a reduced timestep of $dt \sim 6 \times 10^{-6}$~yrs, three orders of magnitude smaller than that used in the original simulations.
At this resolution, even the fastest moving asteroid travels a fraction of a Roche radius in a single timestep and thus a Roche radius crossing is resolvable. 
We carry out an integration with this smaller timestep for $\sim 2$~Myr and remove any bodies which cross the Roche radius from the simulation. 

\subsection{Ejections from the system} \label{subsec:sim_max_exit}
We consider a body to truly have been ejected from the planetary system if it reaches a distance from the central star which exceeds the Hill surface of the system in the Galactic tidal field. 
\cite{Veras2013a} show that this Hill surface is an ellipsoid defined by the following
\begin{equation}
    \label{eq:hills_ellip}
    r_\text{Hill, sys} = \left( \frac{G M_\text{WD}}{\alpha}\right)^{1/3} \mathbfit{k},
\end{equation}
with
\begin{equation}
    \label{eq:hillsellip_k}
    \mathbfit{k} = \left(1, \frac{2}{3}, \frac{ \left[ Q(1+\sqrt{1 + Q} \right] ^{2/3} - Q}{\left[ Q(1+\sqrt{1+Q} \right] ^{1/3}} \right)
\end{equation}
and 
\begin{equation}
    \label{eq:hillsQ}
    Q \equiv - \frac{\alpha}{\Upsilon_\text{zz}}.
\end{equation}
The parameter $\alpha \equiv 4A(A-B)$ relies on the Oort constants whose values at the solar radius are $A = 14.5$~\kmskpc{} and $B = -12.9$~\kmskpc.
The contribution from the disc to the perturbation due to the Galactic tide is represented by $\Upsilon_\text{zz}$ and given by
\begin{equation}
    \label{eq:upsilonzz}
    \Upsilon_\text{zz} = - \left[ 4 \pi G \rho_\text{tot} - 2 \delta \Omega_\text{G}^2 \right],
\end{equation}
where $\rho_\text{tot}$ is the total Galactic density, $\Omega_\text{G}$ is the circular frequency of the star around the Galactic centre and $\delta \equiv -(A-B)/(A+B)$ is the logarithmic gradient of the Galactic rotation curve in terms of the Oort constants.

\cite{Veras2014a} notes that the third term in equation~\ref{eq:hillsellip_k} is always in the range $0-2/3$. Thus, if we take its maximal value and assume our simulated planetary systems are hosted by a $0.6$~M$_\odot$ white dwarf at the solar location in the galaxy, then the system's Hill ellipsoid has dimensions of $r_\text{Hill, sys} \sim \left( 240000, 160000, 160000 \right)$~au.
Thus we consider a body to be truly ejected from their planetary system if they exceed an instantaneous distance from the central white dwarf of $240000$~au.

\subsection{Binary Dissociation} \label{subsec:bin_diss}
The Hill radius for a binary asteroid system can describe the region around a primary asteroid within which a secondary asteroid would orbit the primary despite the gravitational influence of a central star and is given by
\begin{equation}
    r_\text{H} = a_\text{a} \left( \frac{2 M_\text{a}}{3 M_\text{WD}} \right)^{1/3},
    \label{eq:HillRad}
\end{equation}
where $a_\text{a}$ is the circumstellar semi-major axis and $M_\text{a}$, $M_\text{WD}$ are the masses of the individual binary components and white dwarf respectively \citep{Donnison2011}.
Thus if the distance between the binary component exceeds $r_\text{H}$, the secondary should only move under the influence of the central star. 

An alternative condition for a binary system to be considered unbound is if the total energy of the system is greater than 0 \citep{Parker2010, delaFuenteMarcos2017}.
The total energy of a binary asteroid system using a two body approximation can be given by 
\begin{equation}
    E = - \left( \frac{2 G M_\text{a}}{2 a_\text{B}} + \frac{G M_\text{WD} \mu}{2 a_\text{a}} \right),
    \label{eq:totalsysenergy}
\end{equation}
where $\mu = 2 m_\text{a}$ and $a_\text{B}$ is the binary semi-major axis \citep{Donnison2011}.

In this work we consider a binary to be dissociated if the instantaneous distance between the binary components exceeds $R_\text{H}$, and confirm that in all such cases $E > 0$. 

\section{Results} \label{sec:results}

\subsection{A Solar System Analogue} \label{subsec:SSAnalog}
The first planetary system architecture we consider is a Solar System analogue which contains the four giant planets: Jupiter, Saturn, Uranus and Neptune. 
We doubled the semi-major axes of the giant planets to mimic the expected $\approx 50$ per cent mass loss for a solar analogue host star \citep{Veras2020b}.
We gave the giant planets zero initial eccentricity and a small random inclination between $0-1^\circ$.

The Solar System CCKBO region extends between $42$ and $47$~au \citep{Nesvorny2019} and is expected to remain stable at least until the end of the main sequence as seen in figure 3 of \cite{Bonsor2011}.
Thus we populated the expanded, post-main-sequence CCKBO region of $84-94$~au with 100 binary asteroid systems as described in Section~\ref{subsec:setup}.

We then integrate the motion of the white dwarf Solar System analogue for $10^{9}$~yrs, removing the bodies from the simulation if they either enter the white dwarf Roche limit and tidally disrupt or if they exceed the limits of the system's Hill ellipsoid and are hence ejected. 

\begin{figure}
    \centering
	\includegraphics[width = \columnwidth]{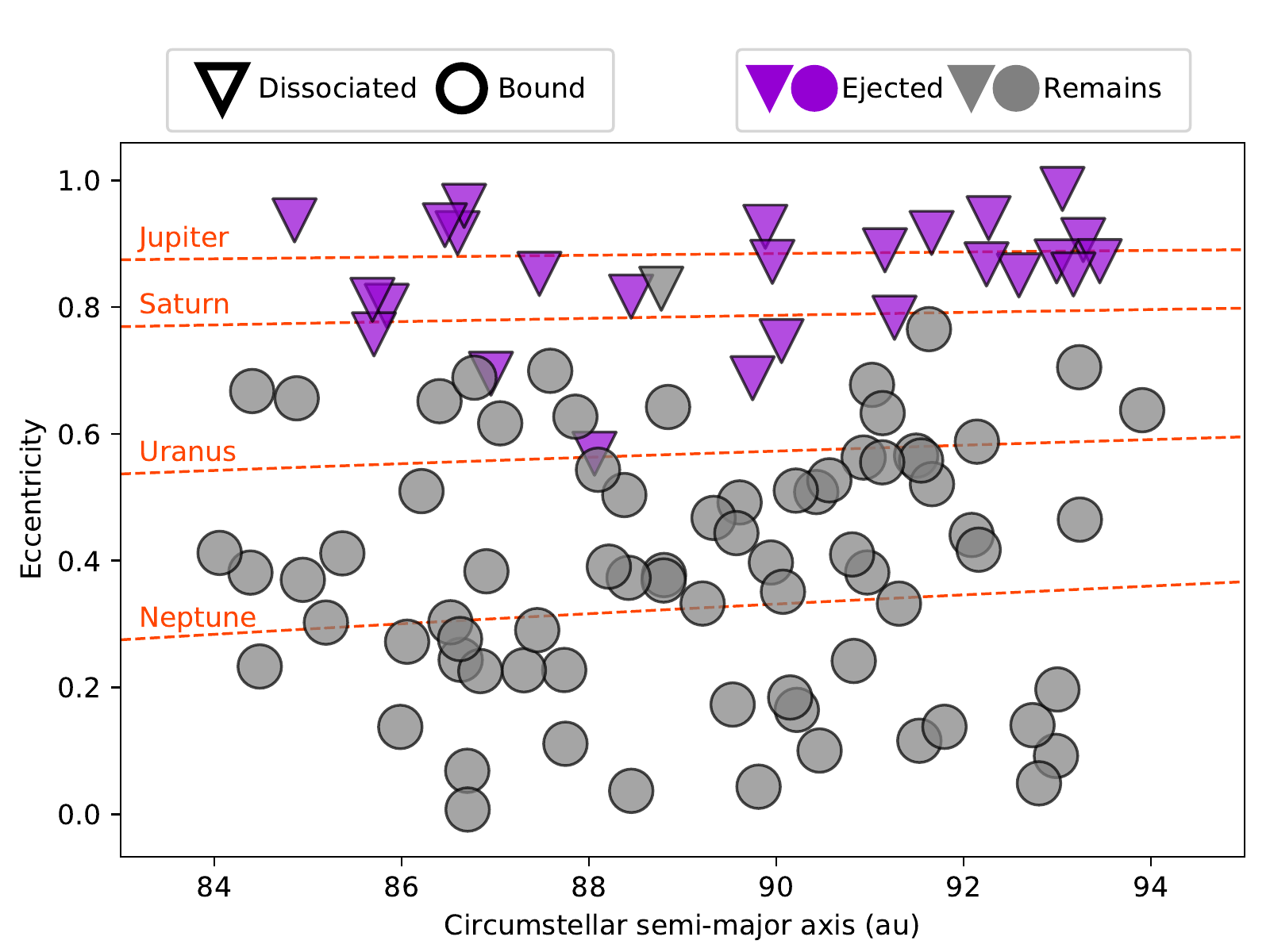}
    \caption{The initial semi-major axis and eccentricity values for the 100 simulated binary asteroid systems in our Solar System analogue. 
    The marker design highlights the final state of the binary; a triangle indicates that the binary dissociates, while a circle shows the binary remains gravitationally bound. 
    Further, the colour of the marker shows the end location of the binary; purple indicates at least one of the components is ejected from the system while grey shows that the binary remains in a stable orbit around the white dwarf.
    The one grey triangle at about $89$~au indicates the only dissociated binary which has remained bound to the white dwarf.
    The dashed orange lines indicate the eccentricity required for a planetesimal to reach a pericentre at the orbital location of the included giant planets as indicated by the annotations above each line.}
    \label{fig:JSUN_ae}
\end{figure}

Fig.~\ref{fig:JSUN_ae} shows the initial semi-major axis and eccentricity values for the 100 binaries considered in our Solar System analogue. 
In this figure, the marker shape identifies the fate of the binary as a whole: triangles indicate that at some point during the Gyr simulation, the binary dissociates as the separation between the bodies is greater than the Hill radius as in equation~\ref{eq:HillRad}, whereas a circle indicates that the binary remains bound.
Further, the colour indicates the final outcome for at least one of the components: purple shows that at least one of the binary components is ejected, while grey shows that both components remain in the planetary system.

From Fig.~\ref{fig:JSUN_ae} it can broadly be seen that higher eccentricity binaries across all semi-major axes are nearly always dissociated and both components are ejected from the system; this appears to hold true regardless of the initial separation of the binary, as is further discussed in Section~\ref{subsubsec:JSUNbinsep}.
There is one case where the binary dissociates according to our Hill radius condition and energy conditions (but does not gain a binary eccentricity larger than unity) and remains within the extent of the white dwarf's Hills ellipsoid at the end of the $1$~Gyr simulation time. 
However, by examining the trajectory of the body across the simulation (see Section~\ref{subsubsec:JSUNejections}), the bodies are clearly undergoing a process whereby their semi-major axes and eccentricities are gradually being increased by successive pericentre passes, and the binary would likely be fully ejected soon after the end of the simulated time. 

The dashed orange lines in Fig.~\ref{fig:JSUN_ae} show the eccentricity values required for bodies across the included semi-major axis range to reach a pericentre value at the indicated giant planet's semi-major axis, thus highlighting which bodies are expected to undergo orbit crossings based on their initial orbits.
It can be seen that all binary systems that cross Jupiter's orbit are both dissociated and ejected from the planetary system.
All bar one system which crosses Saturn's orbit (see Section~\ref{subsubsec:JSUNejections} are again dissociated and ejected.
Although there are a few systems which are dissociated and ejected after crossing Uranus' orbit, largely the ice giants are much less efficient at dissociating and ejecting binary asteroid systems.

\subsubsection{Evolution of binary separation} \label{subsubsec:JSUNbinsep}
\begin{figure}
	\includegraphics[width=\columnwidth]{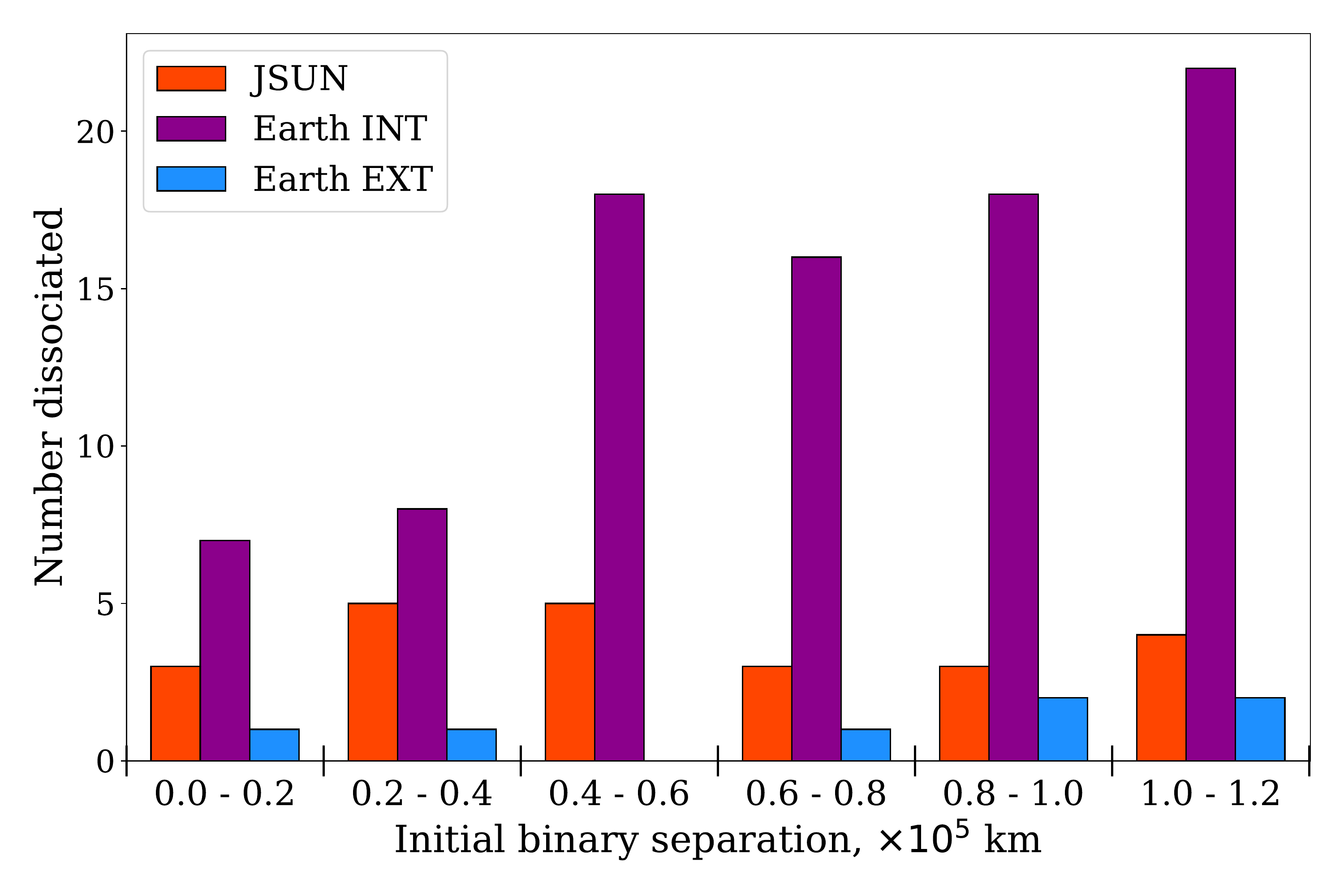} 
    \caption{A histogram showing the percentage of binary systems which dissociate as a function of initial binary separation for all planetary system architectures considered in this work, where the colours indicate the architecture as described in the legend. 
    Here, JSUN refers to our Solar System analogue simulations (Section~\ref{subsec:SSAnalog}), Earth INT, a system with a single Earth mass planet with binary asteroids on interior orbits,  (Section~\ref{subsec:earthmass_int}) and Earth EXT, a single Earth mass planet with binary asteroids exterior to its orbit (Section~\ref{subsec:earthmass_ext}).
    Each architecture's distributions have the same histogram bins, whose widths are identified by the ranges between the large tick marks on the $x$-axis.}
    \label{fig:sepHist}
\end{figure}

To further investigate the effect of the initial separation of the binary components on the probability for the binary to dissociate, we plot a histogram showing dissociations as a function of initial separation in Fig.~\ref{fig:sepHist} for all of the simulation architectures considered in this work.
As can be seen in this figure, for the Solar System analogue (labelled as JSUN in the figure legend) binary dissociations occur across the entire range of initial separation values, with 27/100 binaries in total dissociating. 
When considering all architectures, there is a slight preference for wider binaries to be dissociated, but dissociations occur regardless of the tightness of the binary.

As discussed in Section~\ref{subsec:bin_diss}, we define a binary as having been dissociated if the instantaneous distance between the binary components exceeds the Hill radius of the binary system.
A binary will be gravitationally unbound by a close encounter with a planet if it approaches at a distance $q$ close enough such that the binary separation is approximately its Hill sphere,
\begin{equation}
    \label{eq:bin_diss_dist}
    q \lesssim a_B \left( \frac{3 M_\text{p}}{M_\text{B}}\right)^{1/3},
\end{equation}
where $a_\text{B}$ is the semi-major axis of the binary orbit, $M_\text{p}$ is the mass of the planet and $M_\text{B}$ is the combined mass of the binary system \citep{Agnor2006}.
\cite{Araujo2018} suggest that encounters within $3\times$ and $10\times$ this closest approach distance are still significant encounters which can impart some change onto the binary orbit. 

To further investigate if any of our dissociated binaries have a close planetary encounter prior to dissociation, we utilise the \texttt{REBOUND} Simulation Archive and and carry out higher resolution simulations focussing on the time when the binary dissociates.
Of all the binaries which dissociate, only seven have some kind of significant encounter with one of the giant planets, and in only one of these cases is an encounter within $3\times$ Neptune's closest approach distance immediately followed by a dissociation event. 
Thus planetary encounters does not appear to be the main driver of binary dissociation in this work. 

Most of the binaries which do not dissociate remain on relatively stable binary and circumstellar orbits. 
However, a small number which undergo circumstellar orbit changes, such as orbit crossings, also undergo a corresponding binary orbit change, largely by the binary orbit widening. 
Further, twelve systems which remain gravitationally bound to each other appear to be captured into an orbit at, or close to, the semi-major axis of one of the planets. 
These systems spend an average of $\sim 0.2$~Gyr at, or close to, the semi-major axis of the the planet often before moving back on to wider orbits. 
In all but one case this occurs around Neptune's orbit, in the other the system bypasses Neptune's orbit and gets temporarily captured by Uranus. 

\subsubsection{Ejections from the system} \label{subsubsec:JSUNejections}
We now investigate specific scenarios which lead to planetesimal ejection. 

The Safronov number \citep{Safronov1972} for a planet can provide an indication of how efficient it will be at causing other bodies to be scattered and can be described in terms of the planet's properties as follows 
\begin{equation}
    \label{eq:Safronov}
    \Theta = \frac{a_p}{R_p} \frac{M_p}{M_\text{WD}},
\end{equation}
where $a_p$ is the semi-major axis of the planet, $R_p$ is the radius of the planet and $M_p$, $M_\text{WD}$ are the mass of the planet and white dwarf respectively.

We calculate the Safronov number for each planet included in all simulations discussed in this paper and present the values in Table~\ref{tab:saf}.
\begin{table}
    \centering
 	\caption{Semi-major axis values and calculated Safronov numbers for the Solar System planet analogues considered in this work.}
 	\label{tab:saf}
 	\begin{tabular}{ccc} 
 		\hline
 		 Planet & $a$~au & $\Theta$\\
 		\hline
 		Jupiter & 10.4 & 35.5 \\
 		Saturn & 19.2 & 23.5 \\
 		Uranus & 38.4 & 16.5 \\
 		Neptune & 60.1 & 31.5 \\
 		Earth & 60.1 & 7.1\\
 		\hline
 	\end{tabular}
\end{table}
A Safronov number larger than unity indicates that a planet will be efficient at scattering bodies out of their planetary system, and larger numbers indicate greater efficiency. 
The calculated values in Table~\ref{tab:saf} confirm that more massive planets are more likely to cause planetesimals to be ejected from the system rather than scattered inwards. 

Thus in our Solar System analogue simulations, Jupiter is the most efficient at ejecting planetesimals from the system, although Neptune's Safronov number is similar. 
The main sequence Safronov number's for the same planets are all significantly lower; for Jupiter around the Sun $\Theta = 10.6$.
Although this value is still significantly larger than unity, it is only $\sim 30$ per cent of its post-main-sequence counterpart, highlighting the increased dynamical instability present in white dwarf planetary systems. 

Just over a quarter of the binaries simulated in our Solar System analogue ($26/100$) are ejected fully from the system, with both components reaching a distance larger than $2.4 \times 10^5$~au from the central white dwarf.
We find that $16$ of these binaries reach a minimum circumstellar distance either inside or very close to Jupiter's semi-major axis, with $9$ of these binaries subsequently being ejected separately. 
A further $9$ ejected binaries cross or approach Saturn's orbit and the final binary reaches a minimum distance of $\sim 30$~au, crossing Uranus' orbit. 
This can also be seen in Fig.~\ref{fig:JSUN_ae} with all purple triangles existing above the dashed orange line denoting the required orbital elements to have a pericentre within Uranus' orbit.
All of which confirms that the more massive and close-in a planet is, the more likely it is to eject a planetesimal from its planetary system.

\begin{figure}
	\includegraphics[width=\columnwidth]{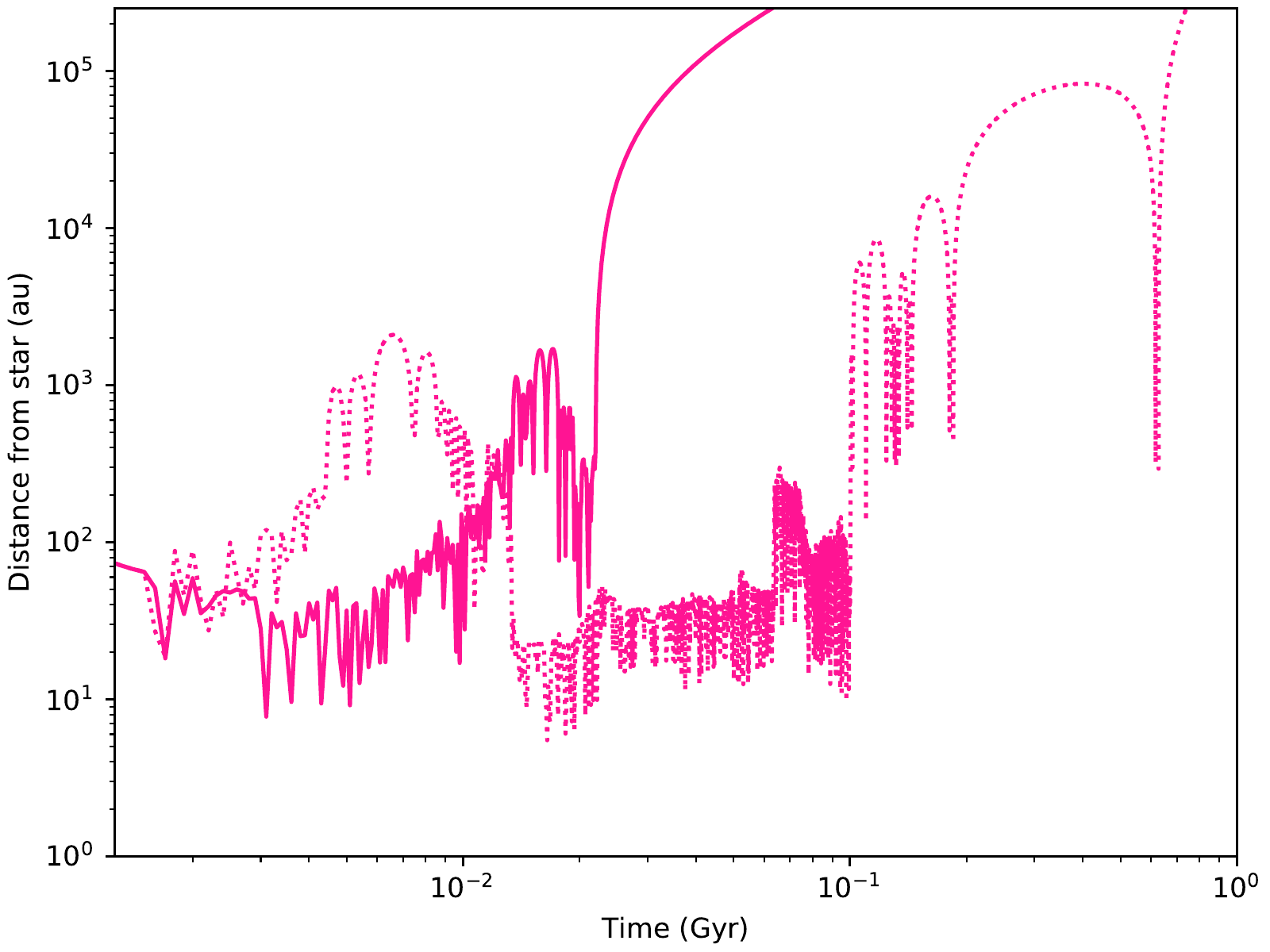} 
    \caption{The ejection process for a single binary asteroid system in the Solar System analogue simulations.
    The plot shows the distance from the central star for the binary primary (solid line) and secondary (dashed line) for the $1$~Gyr simulation time. 
    The binary dissociates at $1.6$~Myr, and the primary is ejected at $54.6$~Myr.
    However, the secondary remains in the system for a further $\sim 660$~Myr, gradually increasing its semi-major axis and eccentricity at successive pericentre passes before finally being ejected from the system. 
    }
    \label{fig:JSUN_eject_log}
\end{figure}
Figure~\ref{fig:JSUN_eject_log} shows the process of ejection for one particular binary system in our Solar System analogue simulations. 
The plot shows that the evolution of the distance between the binary components and the central star, with the solid and dashed lines representing the arbitrarily chosen primary and secondary respectively. 
The binary dissociates $\sim 1.6$~Myr after the beginning of the simulation, and the arbitrarily chosen primary is ejected from the system at $\sim 54.6$~Myr. 
However, the secondary remains in the system for a further $\sim 661.7$~Myr, gradually having its semi-major axis and eccentricity pumped up at repeated pericentre passes before finally being ejected from the system.
10 of the 26 binaries which are ejected here showcase a similar long-term ejection process, seemingly `bouncing' out of the system.

Such increases in semi-major axis and eccentricity are analogous to the processes which populate the Solar System Oort Cloud with comets from the inner planetary system.
Under the gravitational influence of a planet, a comet with semi-major axis much larger than it's pericentre will receive an effective kick in energy at pericentre which leads to increases in semi-major axis and eccentricity with fixed pericentre.
This process leads to a random walk in orbital evolution which has been investigated by many authors \citep[eg][and others]{Duncan1987, Brasser2006}.
Without the influence of galactic tides or stellar flybys to raise the pericentre values of the `comets' outside of the inner planetary system in our simulation, the bodies are ejected.

A similar process for planet mass objects has also been seen in simulations before: the left hand panel of figure 2 of \cite{Veras2009} also shows wide orbits induced by scattering persisting in their planetary systems for hundreds of Myr before being ejected.

Four of these binary systems dissociate according to the condition that the distance between the components exceeds the Hill radius of the binary (equation~\ref{eq:HillRad}). However, their components maintain similar orbits and hence are ejected simultaneously from the system.
As discussed in Section~\ref{subsec:SSAnalog}, one binary system which is dissociated per the Hill radius condition but is not ejected from the system undergoes a similar `bouncing' trajectory to that shown Fig.~\ref{fig:JSUN_eject_log}, where it expected the binary components would be ejected after the end of the simulated time.

\subsubsection{Close approaches to the white dwarf} \label{subsubsec:JSUNapproach}

Directly from our simulation results, we find no instances where a binary component directly crosses the Roche radius of the white dwarf. 
By following the procedure outlined in Section~\ref{subsec:sim_min_exit}, we identify three binary systems which have a component with an expected pericentre below the white dwarf's Roche radius. 
Of those, two are components which are ejected from the system and hence have eccentricities exceeding unity.
The third is not ejected during the course of our original simulations but does undergo a trajectory similar to that displayed in Fig.~\ref{fig:JSUN_eject_log} and hence is expected to be ejected.
We find that no binary component physically reaches the white dwarf Roche radius across the three deeper simulations and hence we find no tidal disruption events across our Solar System analogue simulations.

Although no binary components directly reach the Roche radius of the white dwarf in these simulations, 15 binaries cross Jupiter's orbit and travel further into the inner system.
Of these 15 binaries, nine had initial circumstellar eccentricities $e_\text{i} > 0.9$, five had $0.8 < e_\text{i} < 0.9$ and one had $0.7 < e_\text{i} < 0.8$ implying that an intense dynamical perturbation which can push the body onto a large eccentricity orbit is required for forays into the inner planetary system.
\begin{figure*}
	\includegraphics[width=\textwidth]{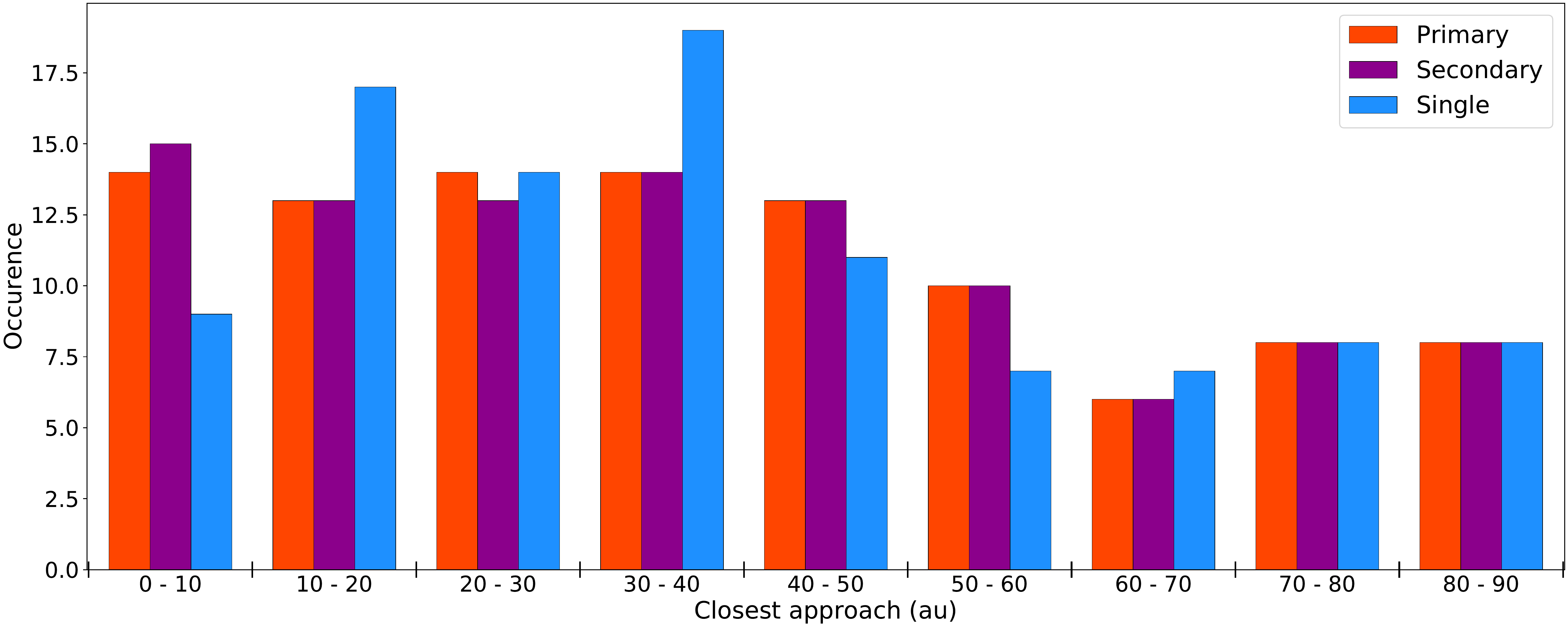} 
    \caption{Histogram of the closest approach distance for each binary component in our simulations; the primary is denoted with orange and the secondary by purple, as indicated in the legend. 
    Broadly the distribution for primary and secondaries are very similar and only diverge at the closest approaches.
    In blue we show the closest approach distance for additional simulations which were run with identical clones of the primary binary component but with the secondary removed.
    In all cases a significant fraction of bodies cross Jupiter's orbit where they may then be further perturbed by remnant terrestrial planets.
    Each distribution has the same histogram bins, whose widths are identified by the ranges between the large tick marks on the $x$-axis.
    }
    \label{fig:JSUN_approach}
\end{figure*}

Fig.~\ref{fig:JSUN_approach} shows a histogram of the minimum distance that each binary asteroid component makes to the white dwarf (orange for the primary and purple for the secondary). 
In this work, as we neglect the presence of any remnant planets, asteroids which cross Jupiter's orbit and enter the inner planetary system ($a \leq 10$~au) may encounter smaller surviving terrestrial planets which could further scatter the planetesimals onto Roche limit crossing orbits.
Overall, three of our binaries approach the white dwarf within $5$~au and a further twelve within $10$~au where they may be more likely to encounter further planets and undergo more perturbations.

\subsubsection{Comparison with single-body evolution} \label{subsubsec:JSUNcompar}

To further understand if the binarity of these systems has had any impact on the evolution of the asteroids, we carried out additional simulations where the secondary component has been removed.
These simulations are begun with otherwise identical initial conditions to the binary simulations through utilising the \texttt{REBOUND} SimulationArchive \citep{ReinTamayo2017}.
Here we discuss some of the results comparing the evolution of a binary asteroid system to its single-body counterpart.

Often, when both components of the binary are ejected from the planetary system, the corresponding single body is also ejected, which suggests the ejection process is tied more to the initial orbital configuration of the system than the binarity. 
However, the timescales for these ejections can differ: in four cases where the binary components are ejected separately, the single body is ejected between those two events.
Further, there are cases where the single can be ejected either before or after the binary components are ejected. 
There are four occurrences where the binary system is ejected but the single body is not and another four where the single is ejected and the binary is not.
For the specific binary asteroid shown in Fig.~\ref{fig:JSUN_eject_log}, the binary primary and secondary are ejected at $\sim 55$ and $\sim 716$~Myr respectively, while the single body counterpart is ejected at $\sim 90$~Myr.

Fig.~\ref{fig:JSUN_approach} shows a histogram of the closest approach for both the binary and single body simulations. 
The distribution of closest approaches into the inner system the binary components reach is shown by orange and purple bars, while the single body counterpart is shown in blue.
For 40 per cent of the simulated systems the difference in closest approach values is less than $1$~au, while another 26 differ in closest approach value by up to $40$~au.
In 54 of our systems, the binary approaches closer to the WD than the single body, in 45 cases the single body makes a deeper entrance to the inner system, and in two cases, the single body arrives between the two binary components. 

\subsection{A single Earth-mass planet with exterior planetesimals} \label{subsec:earthmass_ext}
It has been shown that lower mass planets are more efficient at delivering planetesimals into the inner regions of white dwarf planetary systems where they can undergo further encounters which lead to tidal disruption and eventual accretion \citep[see left hand panel of fig. 6 of][]{Bonsor2011}.
Since our Solar System analogue simulations include the influence of giant planets and show a number of ejected and stable binaries, but no planetesimals on orbits which would suggest eventual accretion, we now change our attention to lower mass planets. 
Figure 6 of \cite{Bonsor2011} shows that in their numerical simulations, the highest number of planetesimals are scattered inwards under the influence of a $1$~M$_\oplus$ planet compared to higher masses ($10-100$~M$_\oplus$). 
In their work, a planetesimal is considered to be scattered inwards if its semi-major axis is less than $a_\text{in}$, taken to be $a_\text{in} = a_\text{p} - 7 r_\text{H}$, where $r_\text{H} = a_\text{p} \left( M_\text{p}/3 M_\text{WD} \right)^{1/3}$.

This value is chosen in their work under the assumption that the planetesimals could then be scattered by remnant planets within the inner system.
Thus, in the context of our work $a_\text{in} = 55.17$~au.

Thus we place a $1$~M$_\oplus$ planet at the WD semi-major axis of Neptune in the previous Solar System analogue ($a = 60.1$~au), again with zero initial eccentricity and a random inclination between $0-1^\circ$.
Then following the procedure set out in \cite{Bonsor2011}, and ignoring the effect of post-main-sequence mass loss on mean motion resonances (MMRs) \citep{Debes2012, Voyatzis2013, Li2021}, we place our binary asteroids from the post-main-sequence orbital location of the planet to the location of the outer 2:1 MMR ($60.1$~au $< a_\text{a} < 95.4$~au).
This 2:1 MMR with Neptune is also the approximate edge of the Solar System's Kuiper Belt \citep{Allen2001, Trujillo2001} and the location of the resonance is defined as 
\begin{equation}
    \label{eq:21mmr}
    a_\text{21MMR} = \left( \frac{2}{1} \right)^{2/3} a_\text{p},
\end{equation}
where $a_\text{p}$ is the semi-major axis of the planet \citep{MurrayDermot1999}.

The 100 binary asteroid systems were again initialised using the same orbital elements as discussed in Section~\ref{subsec:setup}, and then the system was integrated for $1$~Gyr. 

\begin{figure}
	\includegraphics[width=\columnwidth]{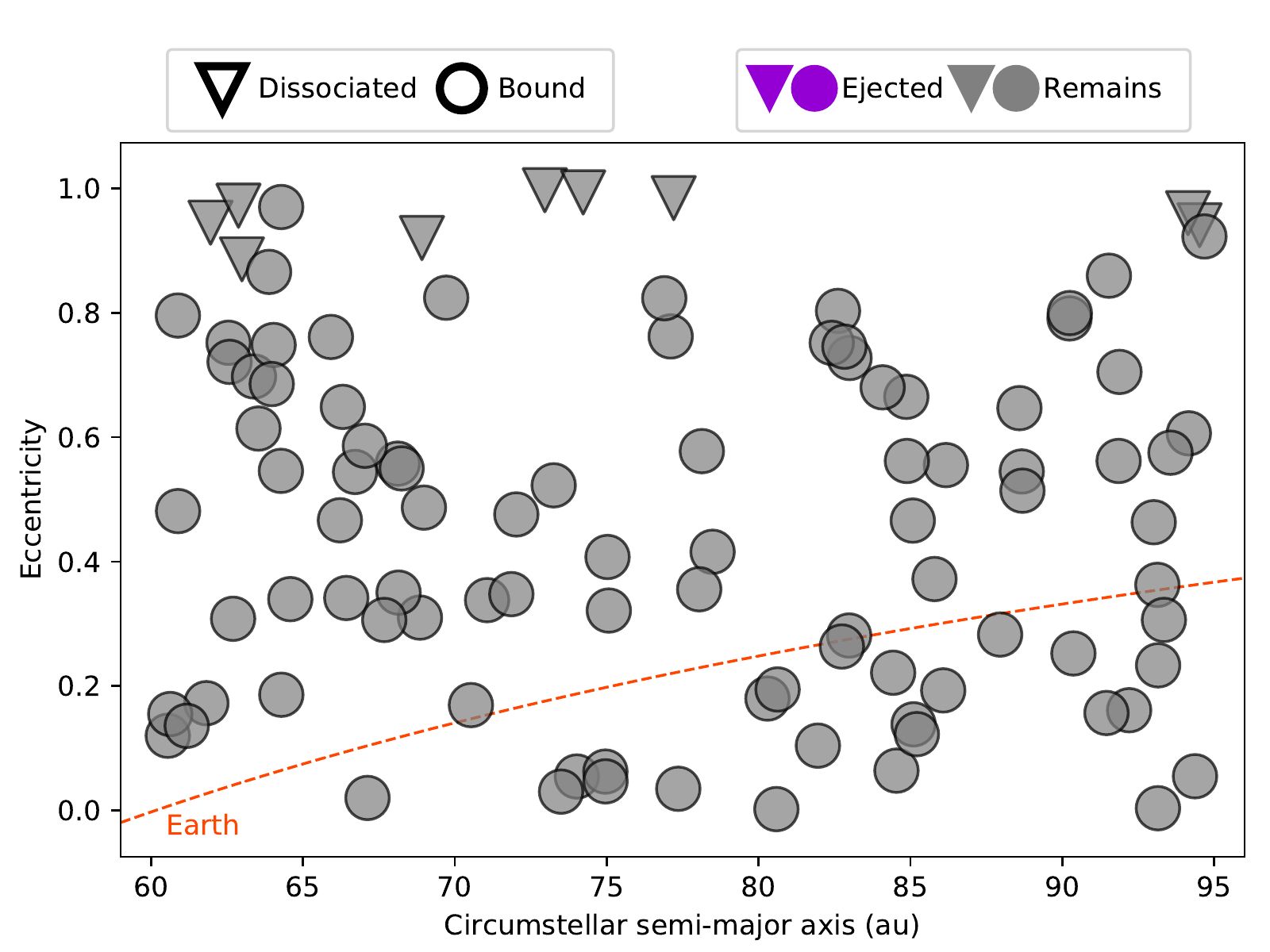}
    \caption{The initial semi-major axis and eccentricity values for 100 binary asteroids on orbits exterior to a single Earth-mass planet. 
    As in Fig.~\ref{fig:JSUN_ae} and as in the legend, the marker shape highlights the outcome for the binary orbit and the marker colour indicates the final state of the stellar orbit. 
    In this case, only binaries with stellar eccentricity $e > 0.9$ dissociate and no binary components are ejected from the planetary system.}
    \label{fig:EarthX_ae}
\end{figure}

Fig.~\ref{fig:EarthX_ae} shows the initial semi-major axes and eccentricities of the binary asteroids considered in these simulations.
The shape markers and colours are indicators of the same as in Fig.~\ref{fig:JSUN_ae} and are described in the figure legend.
From this figure it is clear that all of the binaries remained in the system during the course of the $1$~Gyr integration, with no ejections or bodies being directly scattered to the Roche limit of the star. 
Due to the lower value of the Safronov number for the Earth-mass planet considered here (see Table~\ref{tab:saf}) it is not unsurprising that there are no ejections in these simulations.

Using the definition of the inner planetary system of $a_\text{in} = 55.17$~au from \cite{Bonsor2011}, a quarter of the binaries in our simulation including an Earth mass planet and a quarter from the Solar System analogue encroach in the inner system, seemingly in contrast with the results of \cite{Bonsor2011}.
At closer distances, only two binaries in the Earth mass simulations had a closest approach towards the white dwarf within $1$~au and only another ten binaries approached within $10$~au.
We find no binary components come within $5$ per cent of the white dwarf's Roche radius.
While this amount is a slightly smaller number of bodies entering the deeper regions of the planetary system compared to our Solar System analogue simulations, no bodies in those simulations directly reach within $1$~au of the star.
However, all of the 15 binaries in the Solar System analogue simulations which approach within approximately $10$~au of the white dwarf are subsequently ejected from the system.
Thus, lower mass planets are more efficient at scattering bodies to closer pericentres without ejecting them, so they can stay in the inner system and be perturbed by further remnant planets for longer timescales.

When comparing the closest approaches for the binary systems and their single body counterparts, there is again less difference than for the Solar System analogue, with only 13 systems having a change in closest approach larger than $1$~au and a single single body simulation whose closest approach was $\sim 11$~au closer than the counterpart binary components.

In these simulations a very small number of binary systems (9/100), all of which start out with very high eccentricities ($e > 0.8$), are dissociated, such that the remaining 91 systems stay bound.
The binaries which remain bound do not show as much binary orbital evolution as those in the Solar System analogues, and the few that do undergo widening exhibit the same corresponding change in circumstellar semi-major axis.
These features can be explained due to the smaller gravitational perturbations from a single-Earth like planet than from multiple giant planets.

\subsection{A single Earth-mass planet with interior planetesimals} \label{subsec:earthmass_int}
Having found no tidal disruption events caused by giant planet scattering, we now focus on a particular region of phase space where previous studies have found asteroids to be particularly vulnerable to being scattered onto orbits which lead to tidal disruptions. 

Again we use an Earth-mass planet, but with binary asteroids which now exist interior to the planet's orbit.
\cite{Antoniadou2019} provide scale-free analytical estimates for the outcomes of planetesimals on planar, elliptic periodic orbits under the influence of one planet. 
By examining the middle panel of fig. 7 of \cite{Antoniadou2019}, we identified a particular region of $a$-$e$ space that is predicted to cause asteroids to be perturbed into the inner planetary system and subsequently collide with the white dwarf.

This region led us to model binary asteroids with parameters distributed in the ranges $0.9 < e < 1$ and $0.61 < a_\text{a}/a_\text{p} < 0.64$ where $a_\text{a}$ and $a_\text{p}$ are the circumstellar semi-major axes of the binary system and the planet respectively. 
These values give $36.7 $~au~$\lesssim a_\text{a} \lesssim 38.5$~au assuming the Earth-mass planet is again placed at $\sim 60$~au ($\times 2$ Neptune's current semi-major axis.)

Once again, the binaries were initiated with the same orbital elements as previously discussed and then their motion was integrated for $1$~Gyr. 

\begin{figure}
	\includegraphics[width=\columnwidth]{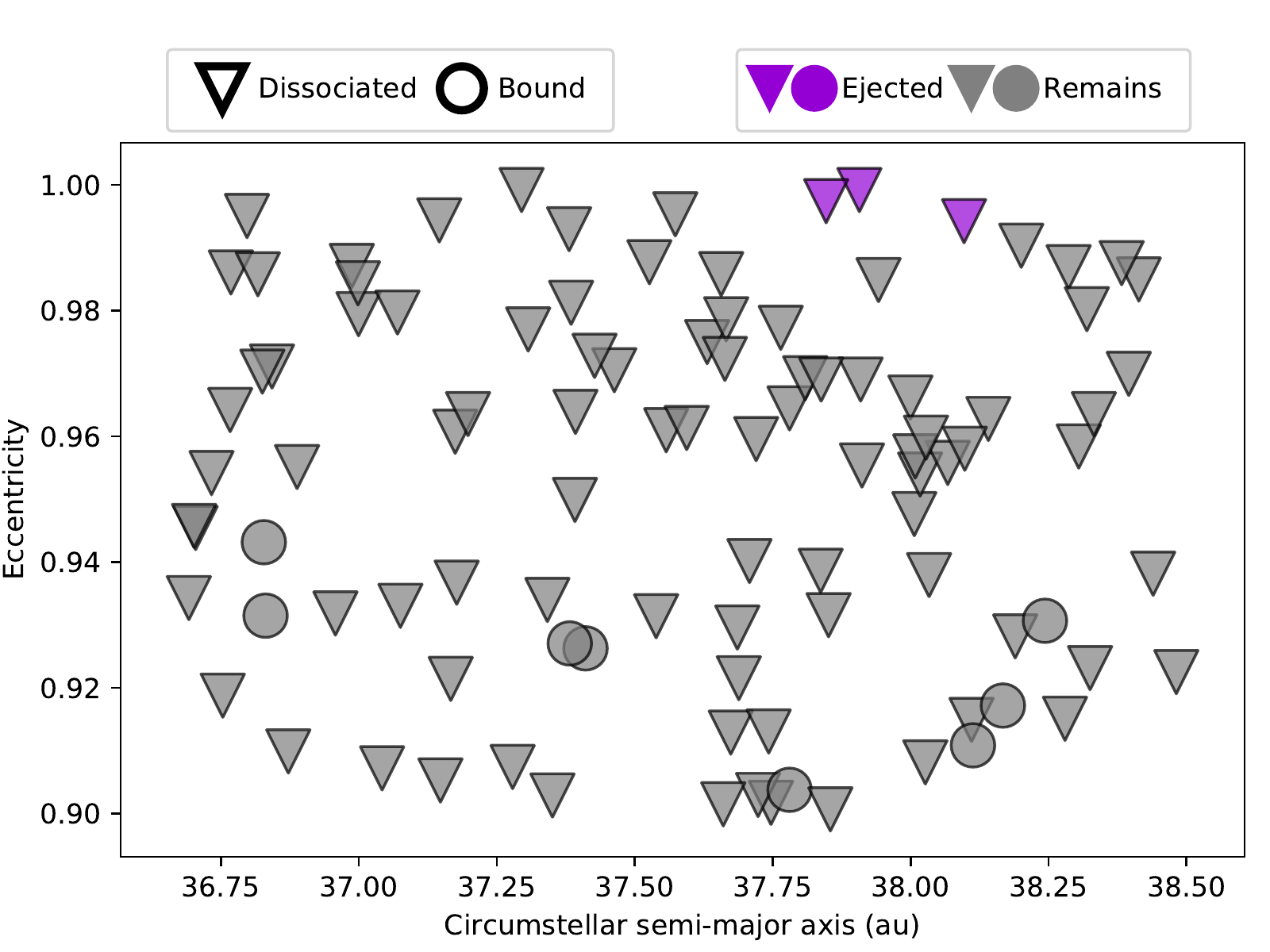}
    \caption{The initial semi-major axis and eccentricity values for 100 binary asteroids chosen to have orbits likely to lead to the planetesimals being accreted onto the white dwarf. 
    As before and the legend, the marker shape highlights the stellar orbit outcome and colour shows the binary orbit end state.
    In contrast to the previous planetary system architectures, here only 8 binaries do not dissociate, all with $e < 0.95$.
    For three separate binaries with $e > 0.99$ and $37.75$~au~$< a < 38.25$~au, they are dissociated and a single binary component is ejected from the system.}
    \label{fig:EarthN_ae}
\end{figure}
Fig.~\ref{fig:EarthN_ae} shows the initial semi-major axes and eccentricities of the circumstellar orbits of the binary asteroid systems. 
The marker styles and colours remain the same as in Figs.~\ref{fig:JSUN_ae} and \ref{fig:EarthX_ae} and as described in the figure legened, but note here the scale on the y-axis which ranges between $0.9 < e < 1.0$ as opposed to $0.0 < e < 1.0$ as before.

As can be seen on this figure, the binaries in this simulation largely have a very different outcome to those planetary system architectures already discussed, with all but eight binaries dissociating before the end of the simulation. 
Further analysis of these dissociation events shows that only 15/92 of the binaries which did dissociate did so after the first $0.5$~Myr of the simulation and none did so after a close encounter with the planet, which may suggest that this region of phase space is mostly inhospitable to binary stability in the first place. 

These features are supported by the fact that the binaries which remained bound all had initial separations smaller than $3.5 \times 10^4$~km, which are some of the tightest binaries we consider in this work. 
Thus, the wider binaries which are typically found in the Solar System Kuiper belt may not be able to exist as binaries at these locations.
The binaries which remained gravitationally bound in this architecture largely also underwent changes in the circumstellar semi-major axis, albeit smaller in magnitude than in the Solar System analogue simulations. 
The corresponding changes in binary semi-major axis were similar in size to those in the Solar System analogue simulations for a similar sized increase in circumstellar semi-major axis. 

As in these simulations we used the same planet mass and semi-major axis as when targetting the exterior 2:1 MMR. The Sarfronov number is again as given in the last row of Table~\ref{tab:saf} (equalling 7.1), which is a sufficiently low value that we should not expect many ejections. 
Indeed, we find only three ejection events, which represents a significant reduction of the level of planetesimals which are scattered to ejection compared to those from the Solar System analogue simulations.

The ejected planetesimals are clustered in the semi-major axis - eccentricity phase space with $37.75 < a_\text{a}< 38.25$ and $e > 0.99$.
In contrast to the ejection events in our Solar System analogue simulations, in these three cases only the arbitrarily chosen secondary of the binary is ejected, with both the primaries and their single body counterparts remaining in the planetary system.
Further, the ejections all occur early in the Gyr simulation at $\sim 1.7$, $4.8$ and $22.6$~Myr, which means that none of the ejections show the same `bouncing' trajectories across hundreds of Myrs as discussed in Section~\ref{subsubsec:JSUNejections}.

The binaries in these simulations were all initiated with smaller semi-major axes and higher initial eccentricities than in the previously discussed architectures, thus the bodies here make much closer approaches towards the white dwarf. 
31/100 of the binaries reach within $1$~au of the central star, and the closest distance to the star that any body achieved was $\sim 0.03$~au, which is still an order of magnitude larger than the Roche radius of the star (see Section~\ref{subsec:sim_min_exit}).
We again find three binary asteroid systems which have a predicted osculating pericentre within $5$ per cent of the white dwarf's Roche radius.
These three systems are those highlighted in Fig.~\ref{fig:EarthN_ae} as being ejected during the course of the simulation and hence with an eccentricity larger than unity.
As outlined in Section~\ref{subsec:sim_min_exit}, we carry out simulations with a smaller timestep centred on the time of the predicted Roche crossing from the osculating orbital elements.
Subsequently we find that no binary component crossing the white dwarf's Roche radius.

There is also less variation in the closest approach distances between the binary and the single body counterparts with only one system where the difference was just larger than $1$~au.

Thus, although we chose to specifically target areas of parameter space which have previously been identified as potential drivers of white dwarf pollution, here we find that further interactions with other planets would be required to cause both binary and single asteroids to reach the white dwarf Roche radius. 

\section{Discussion} \label{sec:discussion}
Although our simulations have not highlighted a particular tendency for binary asteroids reaching orbits which would allow them to be disrupted and then accreted onto the white dwarf over single body counterparts, there are a few particularly interesting consequences of considering binary asteroid evolution. 

\subsection{Implications for white dwarf pollution} \label{subsec:disc_WDpol}
The process of binary dissociation can affect the orbits of the binary components which remain in the system.
All of the binaries which dissociate are ejected in our Solar System analogue simulations, or at least follow trajectories which would lead to them subsequently being ejected. 
Thus, we turn our attention to the effect of binary dissociation on the white dwarf pollution process in our system architectures including Earth-mass planets. 

For the case where the planetesimals begin the simulations exterior to the Earth-mass planet, the final circumstellar semi-major axes of the dissociated primary and secondary can differ by up to $3$~au, while the circumstellar eccentricity differs by only up to $0.01$.
\begin{figure}
    \begin{subfigure}[b]{\columnwidth}
        \centering
        \includegraphics[width = \linewidth]{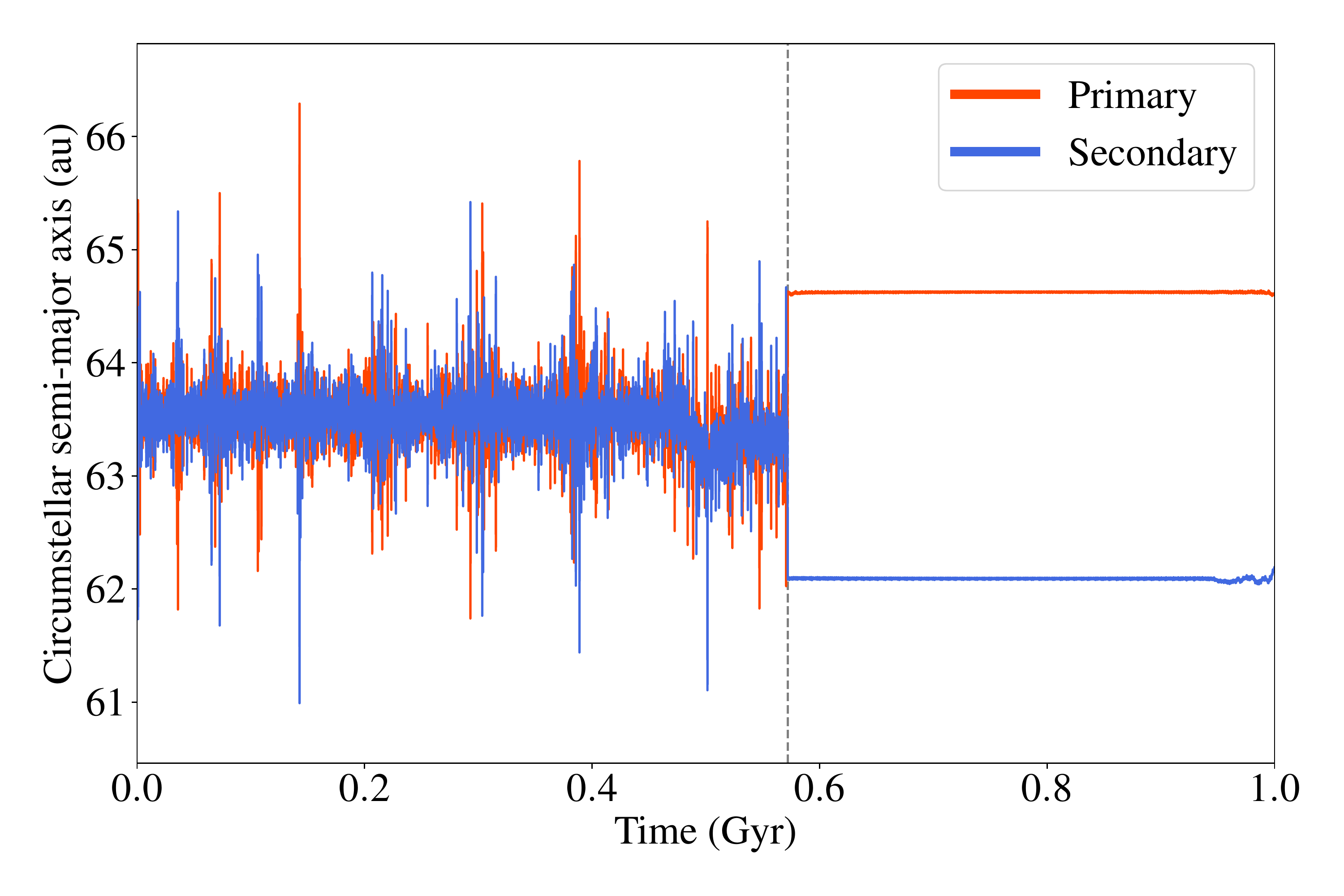} 
    \end{subfigure} \\
    \begin{subfigure}[b]{\columnwidth}
        \centering
        \includegraphics[width = \linewidth]{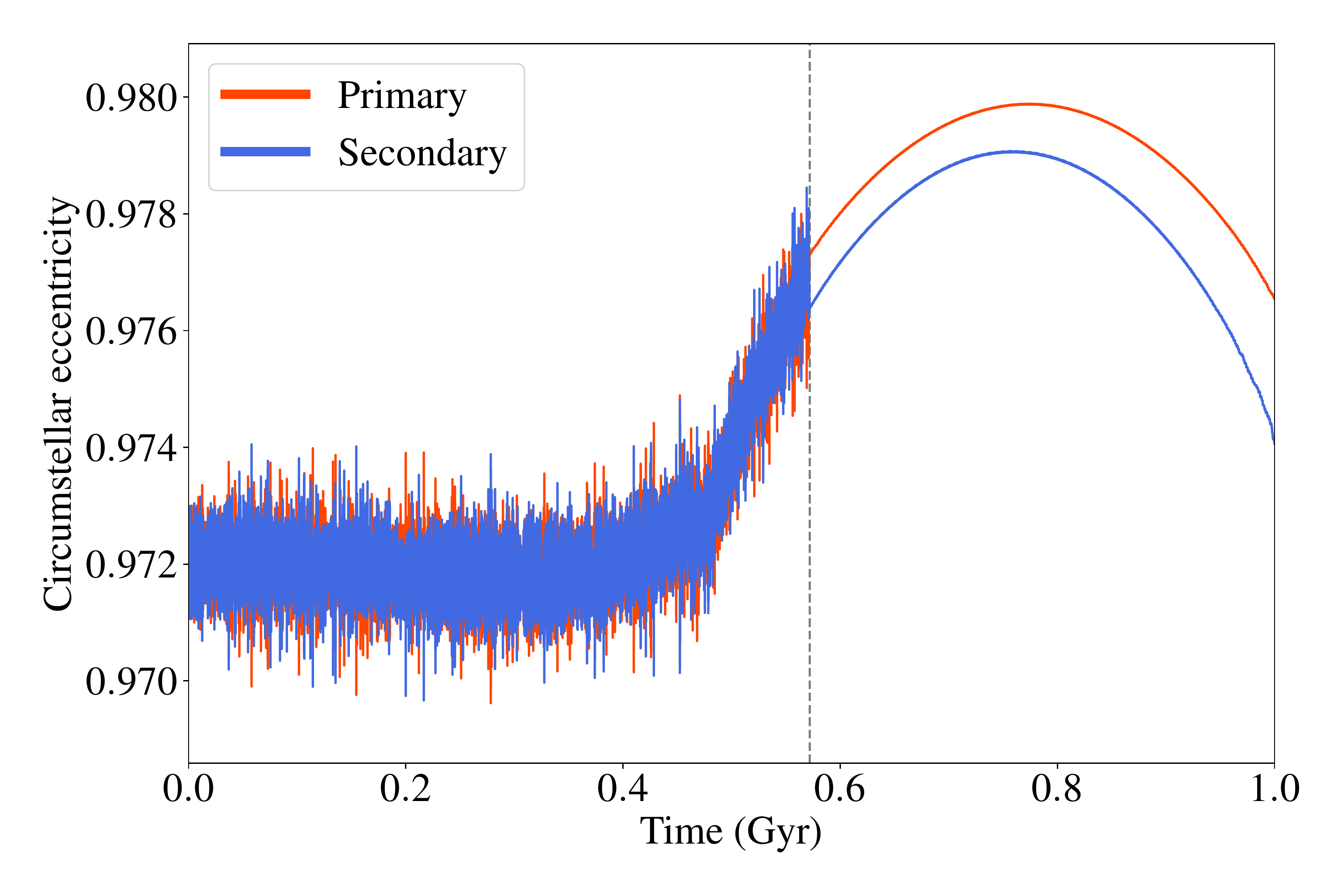} 
    \end{subfigure} \\
    \begin{subfigure}[b]{\columnwidth}
        \centering
        \includegraphics[width = \linewidth]{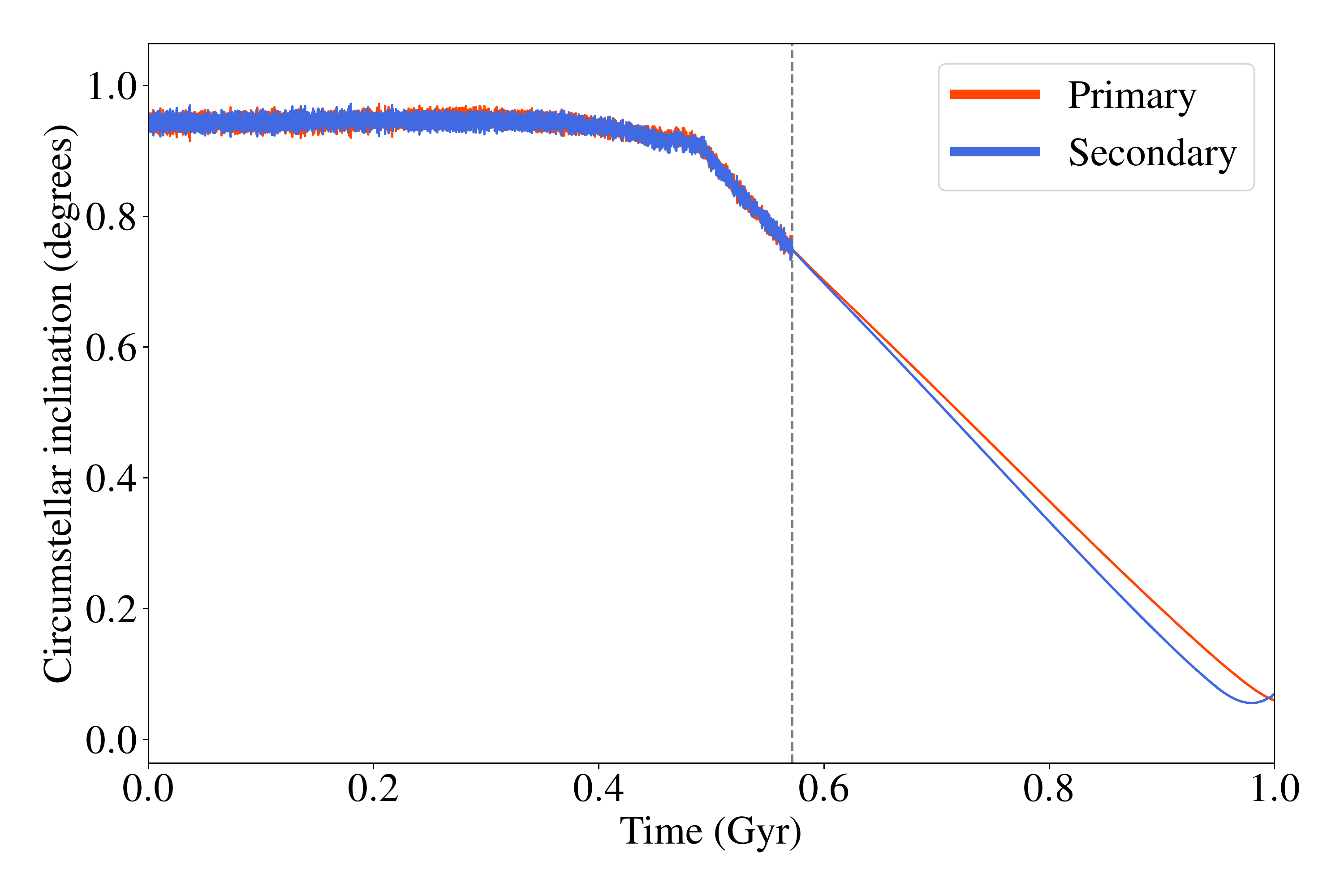} 
    \end{subfigure}
    \caption{The evolution of circumstellar semi-major axis (top panel), circumstellar eccentricity (middle panel) and circumstellar inclination (bottom panel) for the two components of a particular binary which dissociates but is not ejected from an initial orbit exterior to an Earth-mass planet. 
    In each panel, the primary is shown by an orange line, the secondary by a blue line and the vertical dashed grey line pinpoints the time when the binary dissociates. 
    After the point of dissociation the semi-major axes of the components orbits differ by $\sim2.4$~au.
    The circumstellar eccentricity values continue a slow increase that began before the dissociation event, albeit separated by $\sim0.001$, for $\sim1$~Gyr before both begin to decline with an increasing separation. 
    Finally, the inclination values again follow a similar pattern to their pre-dissociation evolution by continuing to decrease from almost $1^\circ$ to $\sim 0^\circ$ by the end of the simulation, with the values for the two components diverging slightly as more time passed from their dissociation. 
    Thus, the dissociation process can affect the subsequent orbital evolution of the components. 
    }
    \label{fig:post_dissoc_evol}
\end{figure}
Fig.~\ref{fig:post_dissoc_evol} shows the circumstellar semi-major axis, eccentricity and inclination evolution for a binary which dissociates from an initial location exterior to an Earth-mass planet and where both components remain in the system for the duration of the simulation.
In Fig.~\ref{fig:post_dissoc_evol}, the coloured lines show the orbital elements of the binary components, as indicated in the legend, calculated assuming they are on Keplerian orbits around the white dwarf and undergoing small perturbations due to the presence of the binary companion.
The vertical grey dashed line shows the time at which the binary is dissociated according to the condition that the instantaneous distance exceeds the system's Hill radius.
In the specific example shown, the binary remains bound for just over $0.5$~Gyr of simulation time, after which there is a gradual increase in eccentricity and decrease in inclination towards the point of dissociation, which occurs at just under $0.6$~Gyr. 

After the dissociation point, the binary primary settles onto an orbit with a semi-major axis $\sim 64.6$~au, while the secondary sits at $a \sim 62.2$~au. 
Thus, the dissociated binary components have a maximum difference in apocentre of $\sim 4.8$~au and $\sim 0.06$~au at pericentre.
The circumstellar eccentricity of the components appear to diverge from each other slightly while still following the same trajectory.
By the end of the simulation, the circumstellar semi-major axis values had been steadily decreasing for $\sim0.2$~Gyr.
Further simulation time, may highlight that the circumstellar eccentricity decreases even further. 
The circumstellar inclinations continue the steady decrease in value that preceded the dissociation event, with both components acquiring $\sim 0^\circ$ by the end of the simulation time.

Thus, the dissociation process can significantly alter the subsequent orbital evolution of the binary components. 
\begin{figure*}
    \begin{subfigure}[c]{\columnwidth}
        \centering
        \includegraphics[width=\linewidth]{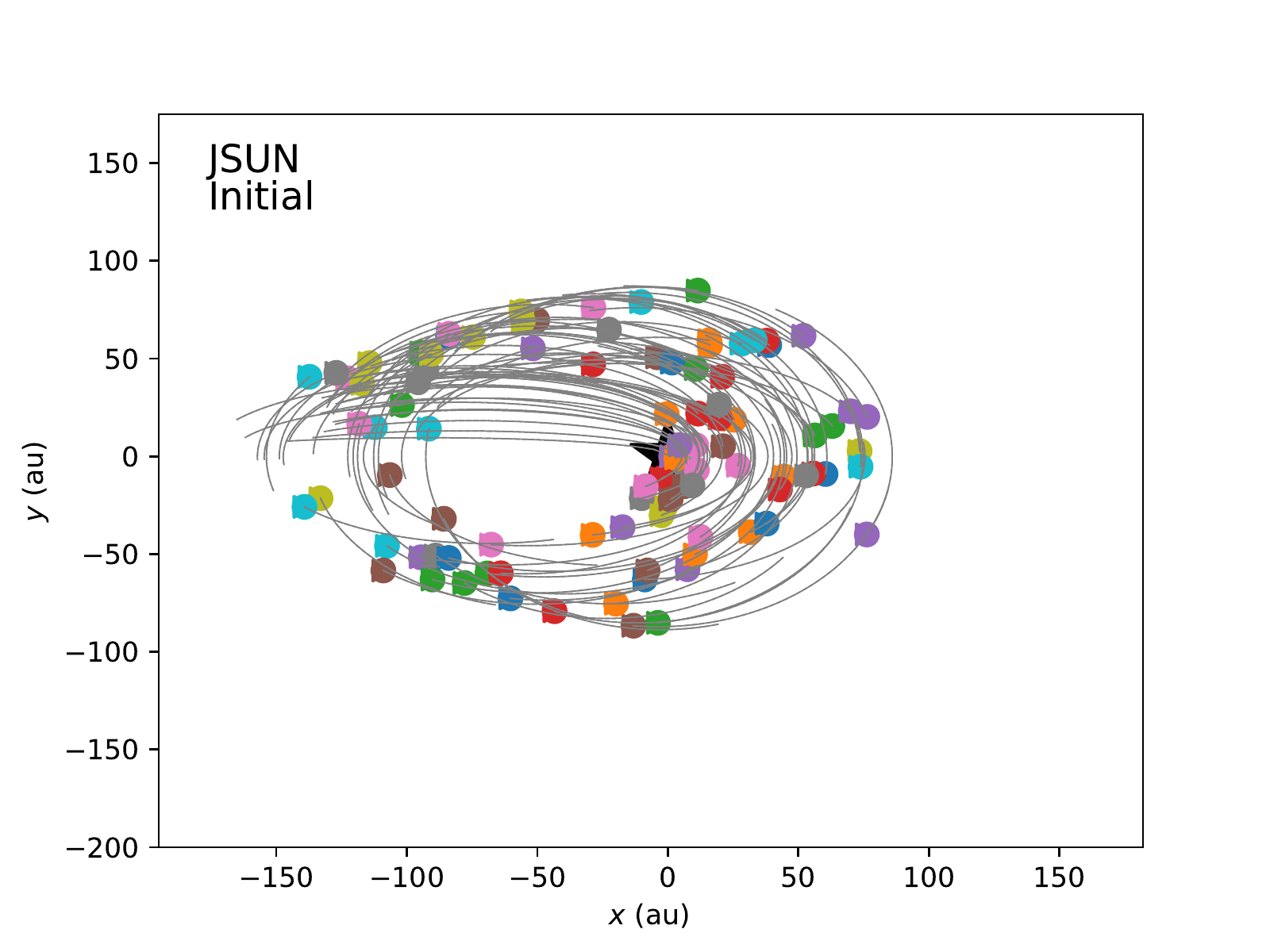} 
        \label{subfig:coolWD}
    \end{subfigure}%
    \begin{subfigure}[c]{\columnwidth}
        \centering
        \includegraphics[width=\linewidth]{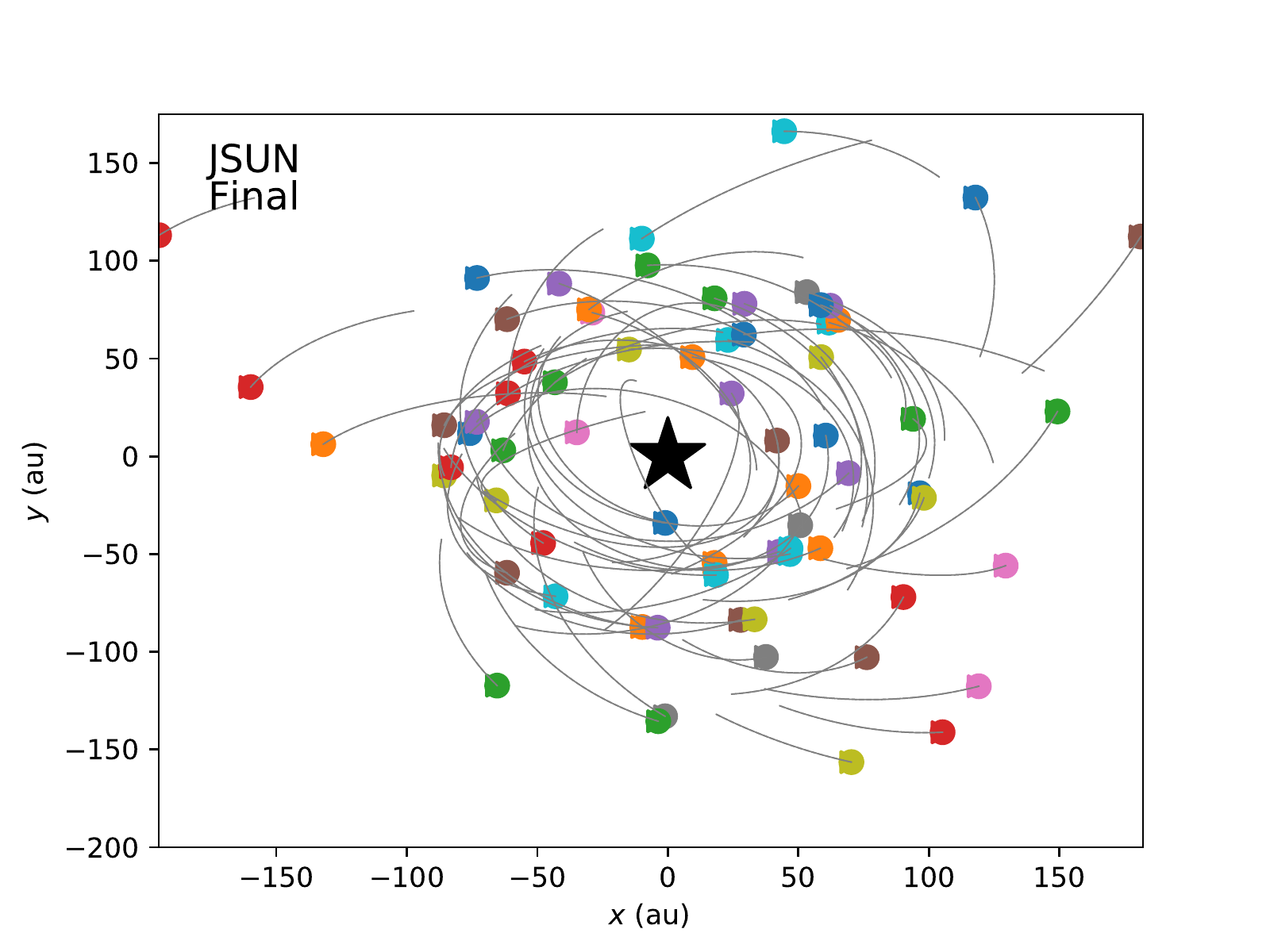} 
        \label{subfig:coolishWD}
    \end{subfigure}
    
    \vspace{-20pt}
    \begin{subfigure}[c]{\columnwidth}
        \centering
        \includegraphics[width=\linewidth]{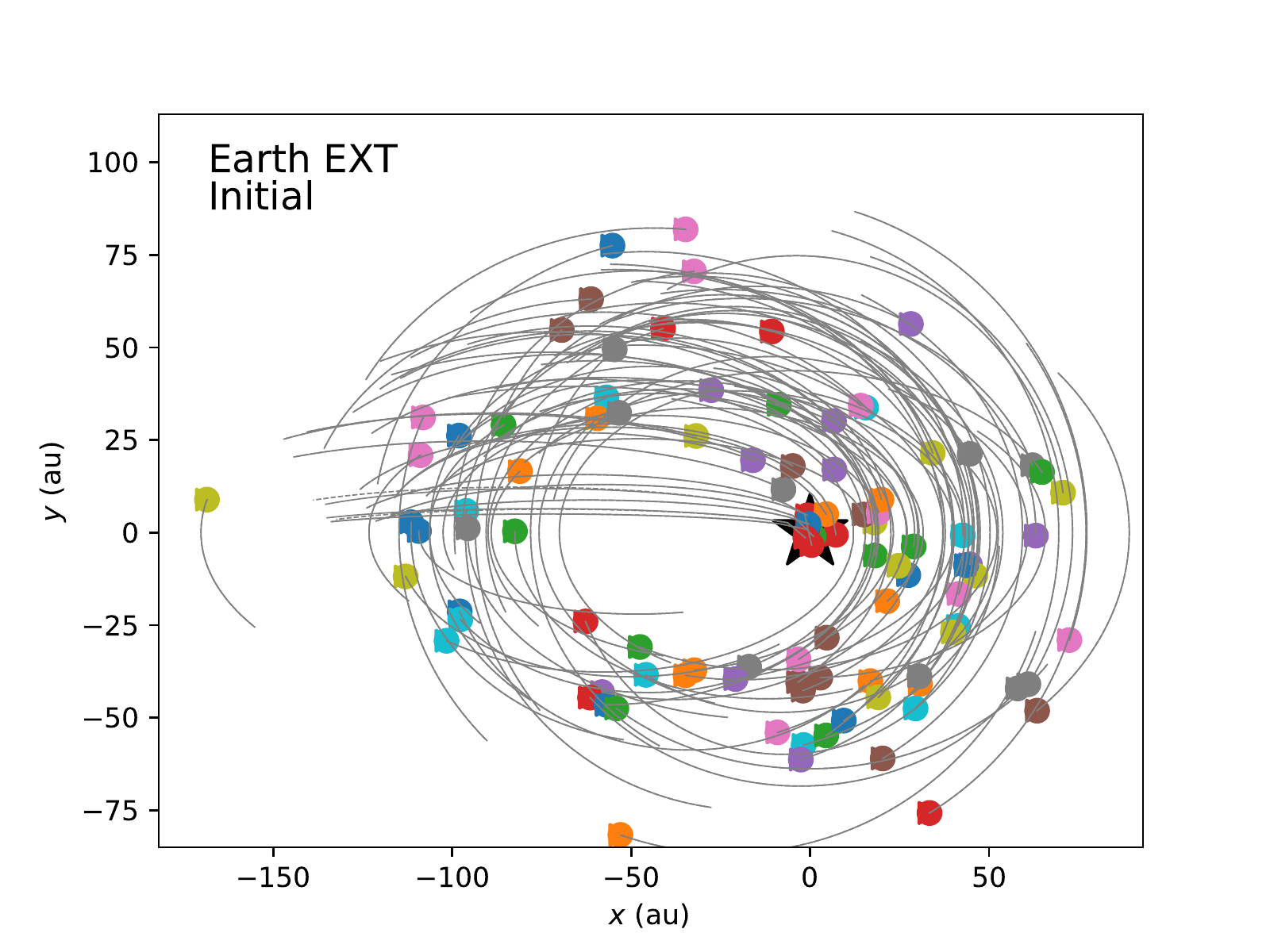} 
        \label{subfig:midWD}
    \end{subfigure}%
    \begin{subfigure}[c]{\columnwidth}
        \centering
        \includegraphics[width=\linewidth]{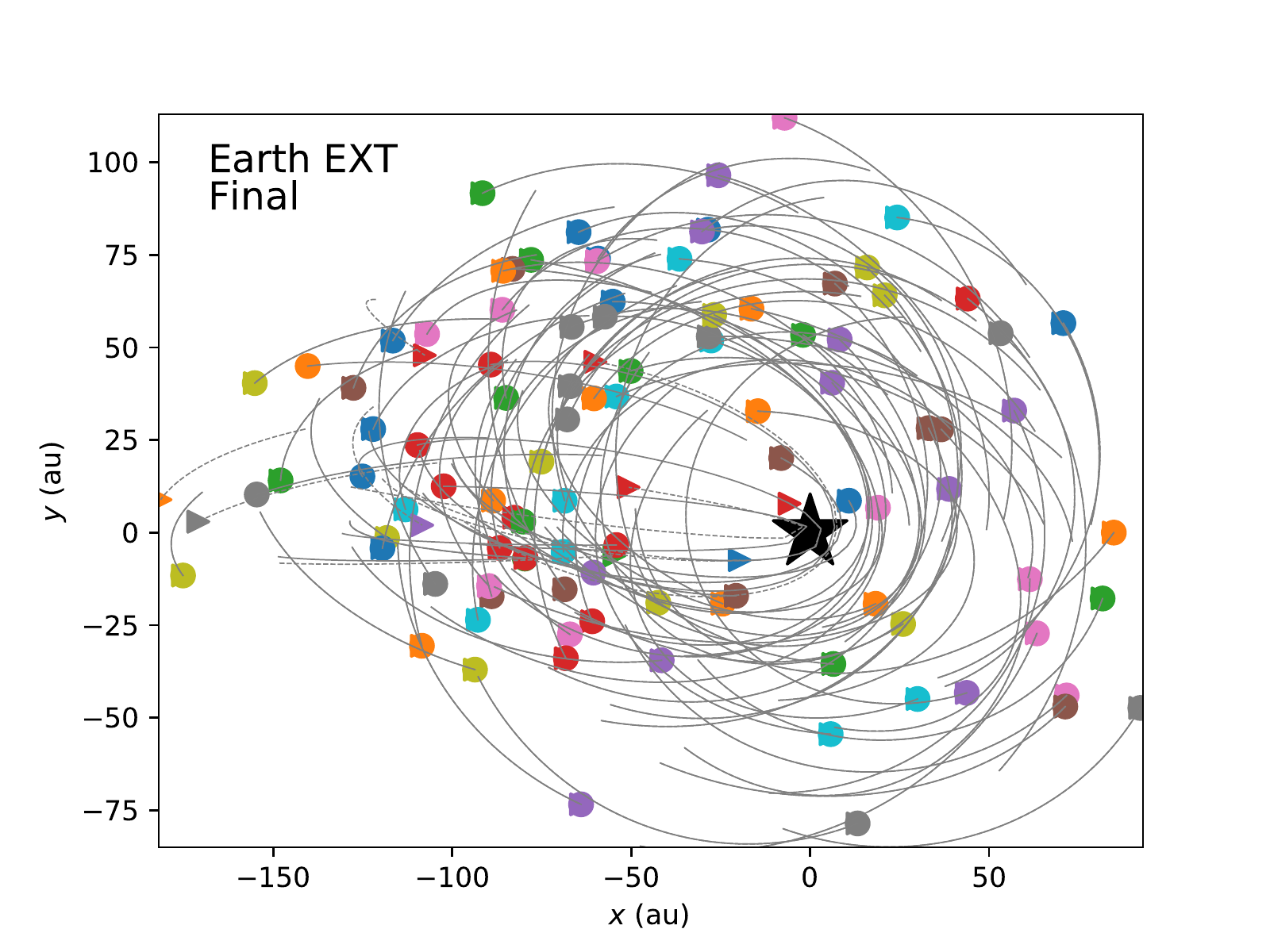} 
        \label{subfig:warmWD}
    \end{subfigure}
    
    \vspace{-20pt}
    \begin{subfigure}[c]{\columnwidth}
        \centering
        \includegraphics[width=\linewidth]{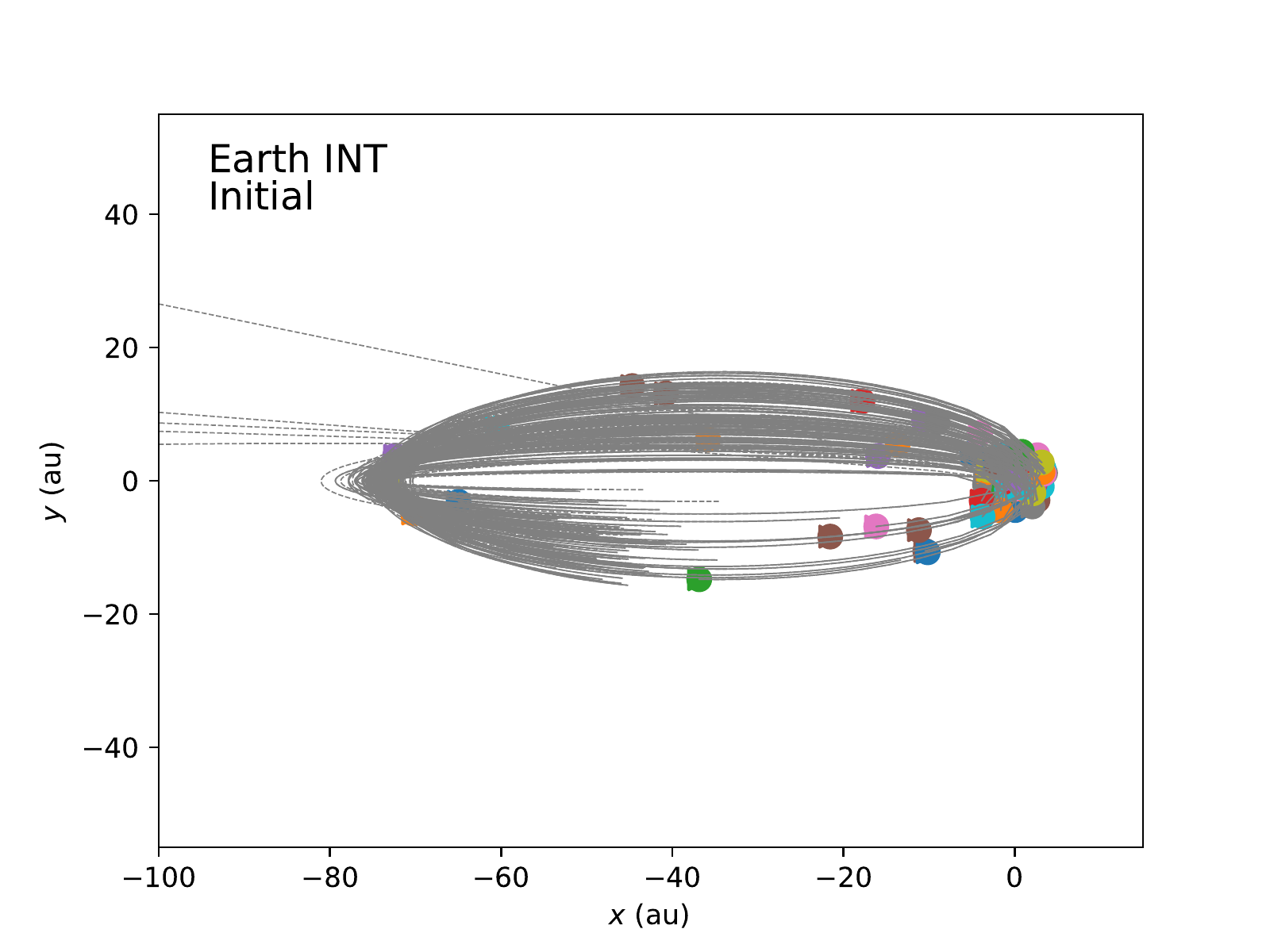} 
        \label{subfig:hotWD}
    \end{subfigure}%
    \begin{subfigure}[c]{\columnwidth}
        \centering
        \includegraphics[width=\linewidth]{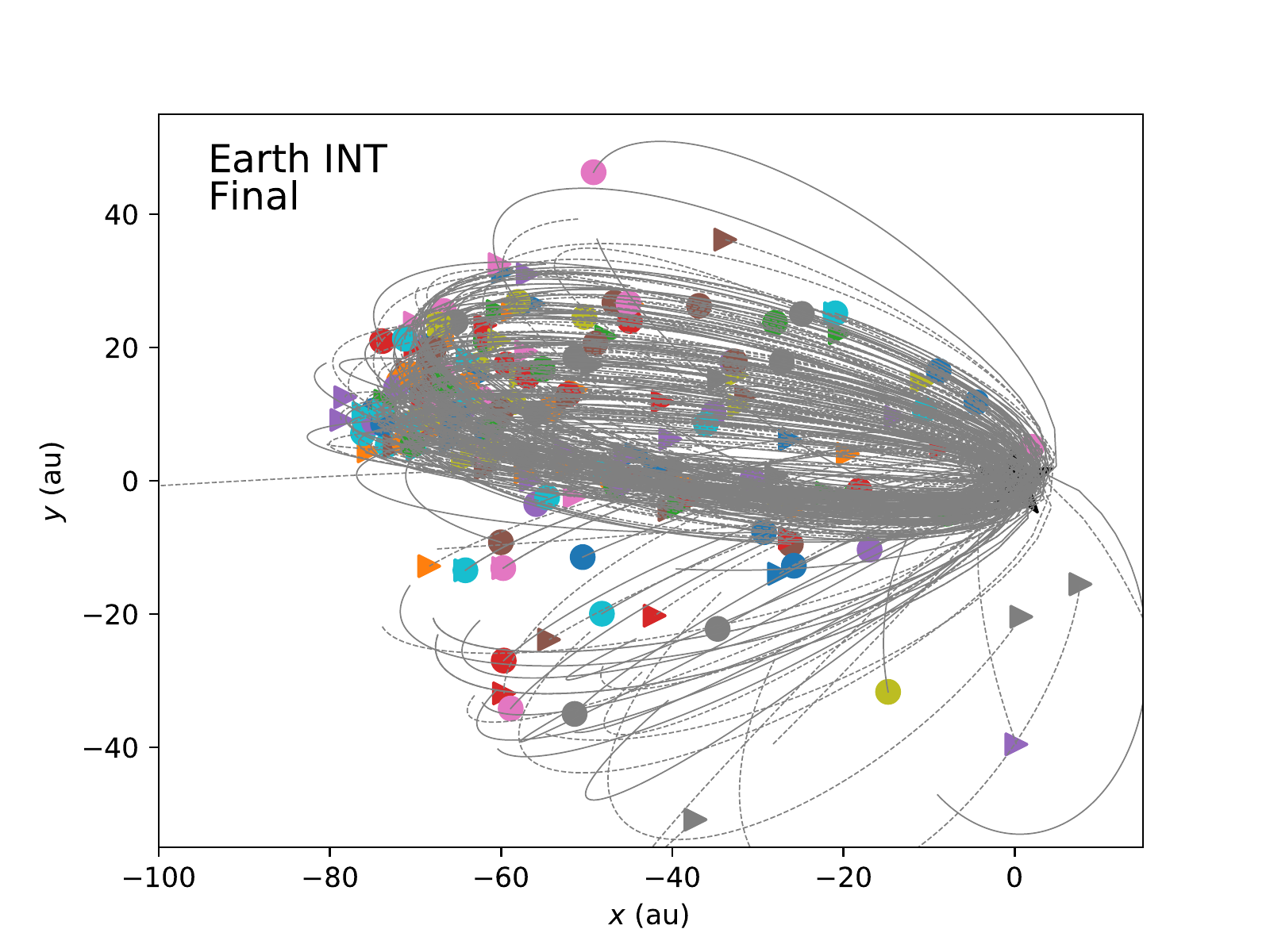} 
        \label{subfig:legend}
    \end{subfigure}%
    \vspace{-18pt}
    \caption{The distribution in $x$-$y$ space of binary components at the beginning of the simulation time (left panels) and then at the end of the $1$~Gyr simulation time (right panels) for all planetary system architectures considered in this work; Solar System analogue (Section~\ref{subsec:SSAnalog}) in top row, binaries exterior to an Earth-mass planet (Section~\ref{subsec:earthmass_ext}) in middle row and binaries interior to an Earth-mass planet (Section~\ref{subsec:earthmass_int}) in the bottom row. 
    The grey lines show the orbital path of the objects for either the following $10^4$~yr (left panels) or the preceding $10^4$~yr (right panels), where the positions have been calculated at higher resolution than in our full simulations, with positions recorded every 100 years. 
    The central white dwarf is located at $(0,0)$ and is denoted by a black star which is hard to distinguish in some panels due to the density of orbit visualisations. 
    The binary primaries are identified by circle markers and triangles of the same colours show the secondaries. 
    By comparing the left and right panels of each row, it is possible to identify how the distribution of planetesimals changes over the course of a Gyr.
    In the top row showing the Solar System analogue results, a number of binary systems have been ejected and those that remain have diffused in the system. 
    In the middle row, dissociated binary components can be distinguished by lone triangle and circle markers.
    No planetesimals are ejected for this planetary architecture but the orbits do become more diffuse by the end of the simulation.
    Finally, in the the bottom row, the most evident feature is the significant evolution of the argument of pericentre which has led to the orbits precessing around the central star.}
    \label{fig:allDistributions}
\end{figure*}
We can examine this effect further by looking at the distribution of binary components both at the beginning and end of our simulations to see how it changes as in Fig.~\ref{fig:allDistributions}. 

Each row shows the initial and final position distributions for the simulated planetesimals in all of our simulated system architectures. 
In the left hand panels we show the initial location of the binaries as coloured markers and the following $10^4$~yrs of their orbital evolution as grey lines, while in the right hand panels we show the final location of the binary components and their preceding $10^4$~yrs of evolution.

The top row of panels show the initial and final distributions of the 100 binary asteroids in our Solar System analogue simulations introduced in Section~\ref{subsec:SSAnalog}. 
The reduction in planetesimal density highlights the $\sim 25$ per cent of planetesimals which are ejected from the planetary system. 
The orbital paths of the binary asteroids appear to have diffused around the star, with a number of bodies being pushed out onto wider orbits.

The initial and final position distributions for binary asteroids on orbits exterior to an Earth-mass planet are shown in the middle panels of Fig.~\ref{fig:allDistributions}.
In this planetary architecture, none of the binaries are ejected from the planetary system and very few are dissociated, but one example can be seen in the lone grey triangle towards the left hand side of the right hand panel. 
The orbital distribution as a whole has also become more diffuse across the simulation time in both the $x$ and $y$ directions.

Finally, the bottom row of Fig.~\ref{fig:allDistributions} shows the position distributions for binaries interior to an Earth-mass planet on orbits expected to cause tidal disruption events. 
By comparing the two panels, it is clear that the orbital distribution of the asteroids changes significantly during the course of the simulation.
Not only can individual dissociated binary components be distinguished, but a number of objects have undergone significant argument of pericentre evolution.

Since these simulations only explicitly considered the gravitational interaction between the planets and the planetesimals, this observed precession is likely to be caused by these gravitational interactions.
The evolution of the argument of pericentre $\omega$, due to the gravitational influence of an external perturber such as a planet in the restricted three body problem can be presented in averaged form as follows 
\begin{equation}
    \label{eq:grav_dwdt}
    \frac{d\omega}{dt} \approx \left( \frac{1}{a_\text{p}^3} \right) \frac{3 G M_\text{p} \sqrt{1 - e_\text{a}^2}}{4 n_\text{a} (1 - e_\text{p}^2 )^{3/2}},
\end{equation}
where $G$ is the universal gravitational constant, $M_\text{p}$ is the mass of the planet, $n$ is the mean motion, $a$ is the circumstellar semi-major axis, $e$ is the eccentricity and the subscripts a and p refer to the asteroid system and planet respctively \citep{Veras2014e}.
For this approximate calculation, we consider the binary asteroid system as a single object whose orbit is approximated by the orbital elements of the primary. 
Taking the range of initial circumstellar semi-major axis and eccentricity values for our simulated binary asteroid systems interior to an Earth-mass planet ($36.7 $~au~$\lesssim a_\text{a} \lesssim 38.5$~au and $0.9 < e < 1.0$), we find an expected rate of change of argument of pericentre of $d\omega/dt \sim 0~-~9^\circ$~Gyr$^{-1}$.
Indeed, we find that a large number of binary systems in our simulations which start with a circumstellar argument of pericentre of $0^\circ$, end the 1 Gyr simulation with $\omega \sim 0-7^\circ$. 
Thus, the observed argument of pericentre evolution is broadly in line with being caused by gravitational interactions with the perturbing planet.

Although we did not explicitly include general relativistic effects in our simulations, these would provide an additional mechanism to induce argument of pericentre evolution. 
\cite{Veras2014d} present the averaged rate of change for the argument of pericentre due to the effects of general relativity as 
\begin{equation}
    \frac{d \omega}{dt} = \frac{3 \left[ G \left( M_\text{WD} + M_\text{a} \right)  \right]^{3/2}}{a_a^{5/2} c^2 \left( 1 - e^2\right)},
\end{equation}
where $G$ is the universal gravitational constant, $a$ is the semi-major axis of the circumstellar orbit, $c$ is the speed of light and $e$ is the orbit's eccentricity.
For the objects in our simulations with binaries interior to an Earth-mass planet, bodies with an initial eccentricity $e = 0.9$ can achieve $d\omega/dt$ on the order of $0.05^\circ$~Gyr$^{-1}$.
But for the very highest eccentricities considered in this work, on the order of $e = 0.999$, $d\omega/dt \sim 5^\circ$~Gyr$^{-1}$.
Considering the argument of pericentre precession due to general relativistic effects would then add an additional rotation to that observed in our simulations due to purely gravitational effects. 

Binaries which do not dissociate and continue to have pericentre passes close to the white dwarf could have interesting consequences for the disruption/accretion processes.
Although as is seen in Section~\ref{subsec:earthmass_int} it can be difficult to have binaries survive on close in orbits, especially if there are large planets in the system, binaries which initially exist exterior to a low mass planet can make close approaches into a planetary system without dissociating. 
How the tidal disruption process might differ if a gravitationally bound binary asteroid system were to cross the Roche limit of their white dwarf is beyond the scope of this paper, but may lead to an interesting distribution of orbiting debris which might be observable through transits. 

\subsection{Implications for interstellar asteroid populations} \label{subsec:interstellar}
The discovery of interstellar objects (ISOs) with the detection of 1I/`Oumuamua in 2017 \citep{Meech2017} and 2I/Borisov in 2019 \citep{Guzik2020} motivated studies into the dynamical processes which lead to planetesimal ejections. 
Alongside ejection during the protoplanetary disk phase \citep{MoroMartin2018}, ejection from a remnant planetesimal reservoir during the post-main-sequence phases of stellar evolution has also been proposed as a mechanism \citep{Rafikov2018, MoroMartin2019, Malamud2020b}.
Although the estimated number density of ISOs originating from post-main-sequence systems is not sufficient to explain the estimated ISO number density from the observation of `Oumuamua, they should still contribute to ISO population. 

\cite{HansenZuckerman2017} predict that as much as $0.1-1.0$M$_\oplus$ of material can be ejected from a white dwarf planetary system hosting Saturn-Jupiter mass planets by considering the average accretion rates for polluted white dwarfs and the fraction of material expected to be accreted versus ejected. 
\cite{Veras2020b} simulated the dynamical evolution of Solar System analogue exo-Kuiper belts consisting of large planetesimals across a star's entire lifetime from stellar cluster birth to the white dwarf phase. 
They find that planetesimals are most vulnerable to system ejections during the stellar cluster phase of evolution, while in the white dwarf phase it is negligible with only a few per cent of planetesimals being ejected (see their figures 4-7).
This predicted lower level of ejections could be due to the planetesimals in their simulations largely remaining on near circular, co-planar orbits, while increasing eccentricities is favourable for ejections in our work. 
The result that there is a negligible difference between the ejections of binary and single component systems in our simulations (Section~\ref{subsubsec:JSUNcompar}) highlights that differences in our orbital distributions is the likely cause for the discrepancies between our results.

In this work we similarly find that giant planets are efficient at scattering material out of their planetary systems and into free floating populations. 
In our Solar System analogue simulations, approximately a quarter of all binary asteroid systems we consider are ejected from their planetary systems when their distance from the central star exceeds the extent of the systems Hills ellipsoid.
The new aspect that our work provides is then not only that there are two bodies to be ejected at every event, but that in $\sim 40$ per cent of our ejection events, the binary components are not ejected simultaneously as discussed in Sections~\ref{subsubsec:JSUNejections}. 
Thus, ejection of a binary asteroid system via planet perturbations around a white dwarf could inject rocky material into interstellar regions across Myrs.

Further, none of our ejected systems are ejected while they remain gravitationally bound as a binary, thus we may not expect a population of free-floating binary asteroids.
Here we aim to expand on this possibility and identify any regions of orbital space where a bound binary may be ejected.
\cite{Jackson2014} present an analytic approach relating the before and after orbit of an object which receives a velocity kick from an external perturbation. 
Although their work focussed on direct impacts between planetary embryos, this is also applicable to planetesimal scattering by planets if we assume that the interaction is impulsive. 
Their formulism allows us to write the minimum kick velocity which leads to a body being ejected from its orbit as 
\begin{equation}
    \Delta v_\text{min} = v_\text{k} \left[ \sqrt{2} \left( \frac{1+e \text{cos}f}{1 - e^2} \right)^{1/2} - \left( \frac{1 + 2e\text{cos}f + e^2}{1- e^2} \right)^{1/2} \right],
    \label{eq:v_ejec}
\end{equation}
where $v_\text{k} = \sqrt{G\left(M+m\right)/a}$ is the circular speed at the orbital distance, $e$ is the eccentricity and $f$ is the true anomaly.
Equation~\ref{eq:v_ejec} allows the identification of the minimum velocity kick required to unbind the binary and circumstellar orbit separately and identify if bound binary asteroid systems can be ejected from their planetary systems. 

First, we turn our attention to identifying how the velocity kick required to unbind the binary changes with the separation between the binary components.
\begin{figure}
	\includegraphics[width=\columnwidth]{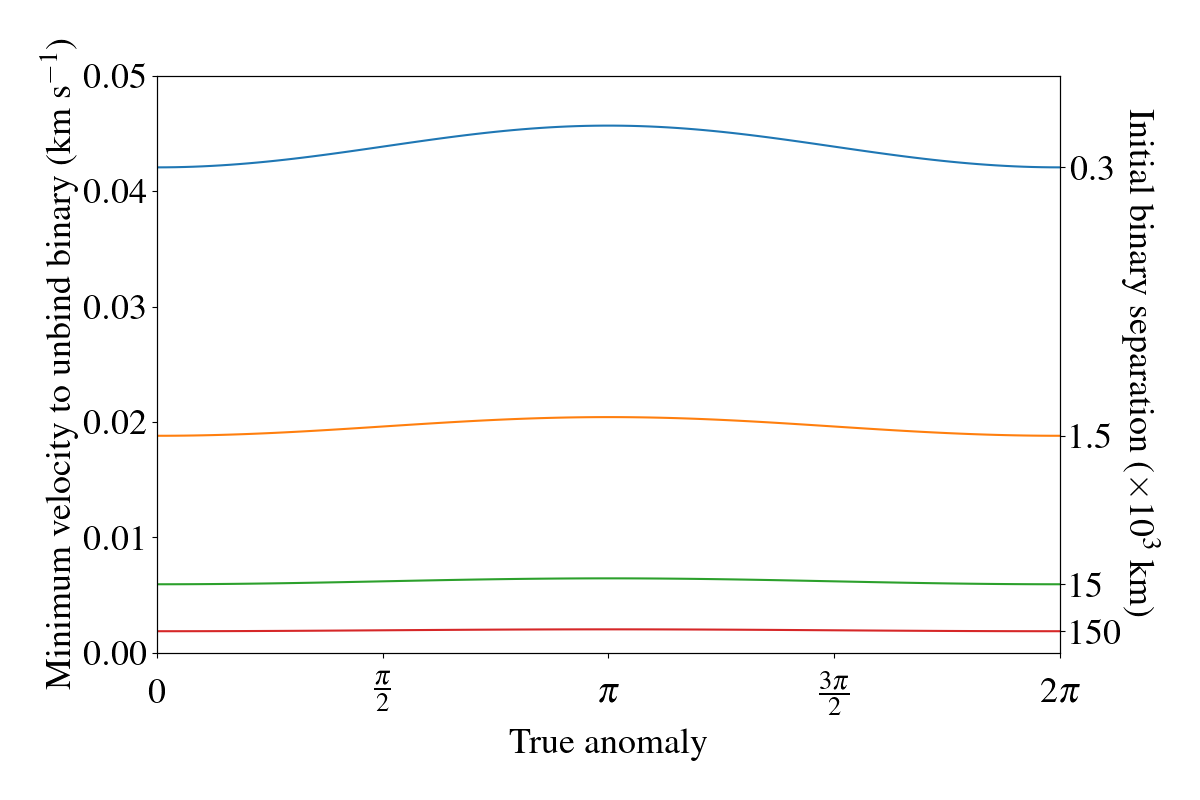} 
    \caption{Minimum velocity impulse required to unbind a binary asteroid system as a function of the binary's true anomaly.
    A range of initial binary semi-major axes are considered and highlighted by labels on the right hand $y$-axis and the different colour curves. 
    The highest required velocity kick and hence the most resilient binary occurs for the tightest binary with a separation of just $300$~km.
    The minimum velocity required peaks at $f = \pi$, but largely does not vary across the orbit.}
    \label{fig:vmin_binary}
\end{figure}
Fig.~\ref{fig:vmin_binary} plots the minimum kick velocity required to unbind the binary as a function of the true anomaly assuming that the binary orbit has a near-zero initial eccentricity as in our simulations. 
We plot the distribution of velocity kick required for initial binary semi-major axis values in the range $300$~km $< a_\text{B} < 1.5 \times 10^5$~km as indicated on the right-hand side of Fig.~\ref{fig:vmin_binary}.
This binary semi-major axis range extends to lower values than those introduced in Section~\ref{subsec:setup} as we aim to identify the tightest binary which might survive, thus we include the semi-major axis of the tightest TNO (a Plutino) identified in fig.~2 of \cite{Nesvorny2019}.

The blue line at $300$~km separation in Fig.~\ref{fig:vmin_binary} is the tightest binary and requires the largest velocity kick to unbind, with the wider binary orbits typical of the CCKBO region needing a much smaller velocity kick.
Although the top most line does show a slight peak at $f = \pi$, the distributions are relatively flat, especially as the binary separation increases, and the peak velocity can approximately be taken as constant across the orbit.
Thus, for the purpose of identifying if it is possible for a binary asteroid system to be ejected while bound, we focus on the most resilient and tight binaries which require a peak $\Delta v_\text{min, B} \sim 0.046$~kms$^{-1}$ in order to be unbound.

For a gravitationally bound binary asteroid to be ejected from its planetary system, the binary must receive an impulsive kick small enough to not unbind the binary but sufficient to unbind the system from the central star, or $\Delta v_\text{min, B} > \Delta v_\text{min, *}$.
If we take the most resilient binary asteroid system identified from Fig.~\ref{fig:vmin_binary} as the most likely binary to survive ejection, we can calculate the above inequality across a range of circumstellar orbits and identify if it is possible to eject a bound binary.

The range of circumstellar semi-major axis values considered in our Solar System analogue simulations ($84$~au $< 94$~au, see Section~\ref{subsec:SSAnalog}) display a narrow range of $\Delta v_\text{min, *}$ for any given eccentricity which all vary by less than $3$ per cent from the average value across that range.
Thus, in the following calculations we use $a_* = 94$~au as a proxy of the entire semi-major axis range.

We then calculate the ratio $\Delta v_\text{min,B}/\Delta v_\text{min,*}$, assuming a fixed $\Delta v_\text{min,B} = 0.046
$~kms~$^{-1}$, across a range of circumstellar eccentricity and true anomaly values and identify regions where the ratio exceeds unity and hence the binary could be ejected while still bound.
\begin{figure}
	\includegraphics[width=\columnwidth]{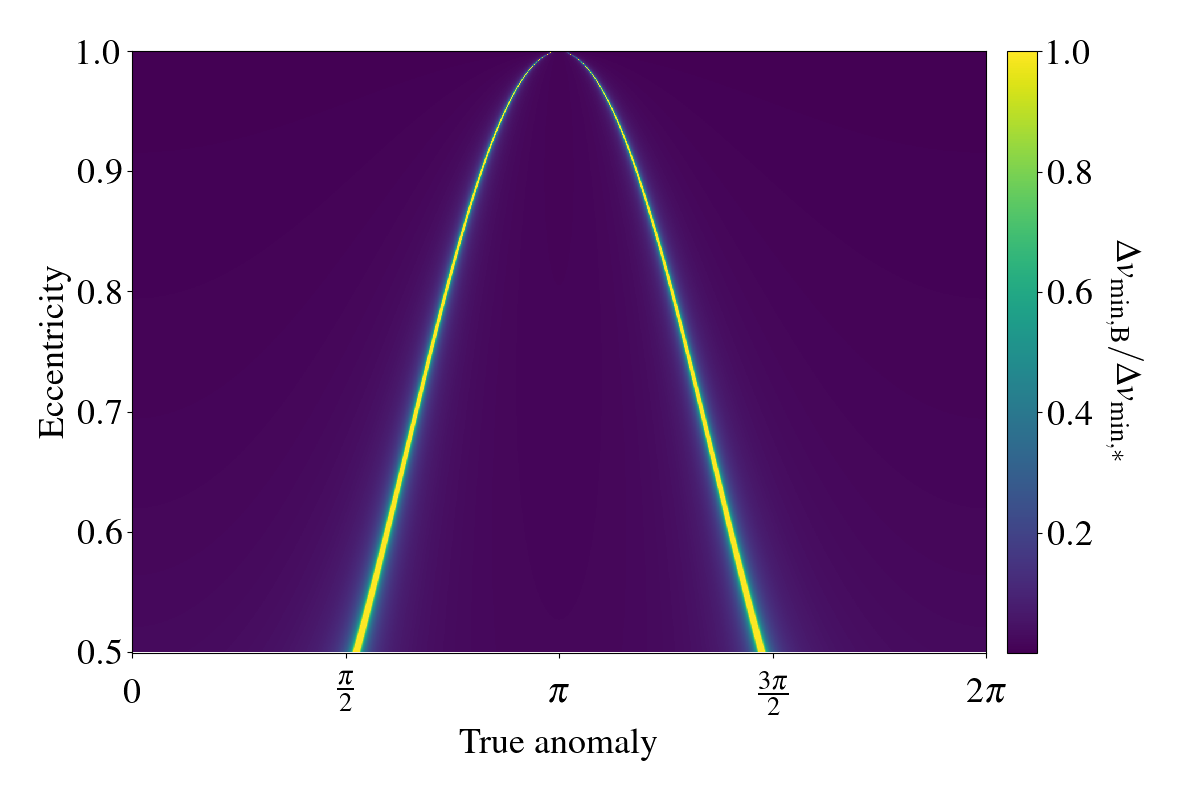} 
	\caption{The ratio $\Delta v_\text{min, B} / \Delta v_\text{min, *}$ for a range of circumstellar eccentricity and true anomaly values. 
	Yellow regions indicate regions of phase space where the above ratio exceeds unity, as demonstrated by the colour bar, and a bound binary could be ejected from its planetary system.
	Across the whole range of eccentricity values sampled, there is at least one location along the orbit where a bound binary could be ejected, although these regions are very narrow.
	Thus, a bound binary could only be ejected if it received a small velocity kick at one of the highlighted regions of it's orbit.
	}
    \label{fig:eject_v}
\end{figure}
Fig.~\ref{fig:eject_v} identifies where the inequality $\Delta v_\text{min, B}/ \Delta v_\text{min, *} > 1$ is true across a range of circumstellar eccentricity and true anomaly values.
As described by the colour bar, yellow regions indicates where the above inequality is true and thus where it could be possible to eject a bound binary.

Although there are two orbital locations at each circumstellar eccentricity value where such an ejection is possible, a very small impulsive kick is required in order to keep the binary gravitationally bound.
Previous studies looking at planetary encounters with binary asteroid systems consider hyperbolic encounter velocities on the order of $\sim 10$~\kms \citep{Fang2012, Meyer2021}, much larger than that required here. 
Thus the coincidence of the small impulsive kick required and the limited region of orbital space for this kick to occur makes the ejection of a bound binary asteroid system unlikely, as seen in the results of our simulations.

As binary asteroids are though to be a primordial consequence of planetesimal  formation and thus should be present in exoplanetary systems, the impact of binarity on ISO production channels could be important to consider in future work. 

\section{Conclusions} \label{sec:conc}
Understanding how we can form the observed debris systems and polluted atmospheres of some white dwarfs is an important aspect of our knowledge of the future evolution of currently observed main-sequence planetary systems as well as the fate of our own Solar System.
While our current knowledge of planetesimal populations around other stars is limited due to observational difficulties, we can use our much better understanding of the Solar System minor bodies to improve our extra-Solar modelling attempts. 

In this work we considered how binary asteroids would evolve during the white dwarf phase of stellar evolution under the presence of different planetary architectures.
This is an important area to study as binary asteroids are so prevalent in the Solar System and likely as prevalent in extrasolar systems, thus they could play an integral role in the evolution of white dwarf planetary systems. 
We carried out computationally demanding N-body simulations using \texttt{REBOUND} to follow the evolution of multiple sets of 100 equal-mass asteroid binaries embedded within different planetary architectures: within a Solar System analogue containing the four giant planets (Section~\ref{subsec:SSAnalog}), and located interior and exterior to a single Earth-mass planet (Sections~\ref{subsec:earthmass_int} and \ref{subsec:earthmass_ext}) respectively. 
We also performed comparable simulations with a single binary component removed, and targeted instability in some simulations by appealing to known periodic orbits.

The first result from our simulations is that higher mass planets are more efficient at ejecting binary planetesimal systems from their planetary systems than lower mass planets, a result already well known for single component systems. 
Around a quarter of the binary systems in our Solar System analogue simulations with giant planets are ejected, but only $1.0$ per cent of binaries are ejected with Earth-mass planets present.
The processes which lead to the ultimate ejection of bodies differ with planet mass, as both binary components are consistently ejected in the Solar System analogue case, while only a single component is ejected by Earth-mass planets. 

We find across all our numerical simulations that no binary asteroid systems are ejected while gravitationally bound.
A further analytic investigation finds that an extremely low velocity impulse from a planetary encounter is required in order to eject a bound binary asteroid system. Thus, we do not expect a population of free-floating binary asteroids.

Lower mass planets are less efficient at dissociating binary systems, unless they are in a region of phase space where single body asteroids are expected to be unstable. 
Binaries which are unbound but not subsequently ejected from their planetary systems evolve to have a broader semi-major axis distribution and also undergo changes in eccentricity and inclination, all of which changes the distribution of objects available to form white dwarf debris systems. 

Thus, while asteroid binarity may not directly affect the production of white dwarf debris, it can help shape the population of planetesimals available to be disrupted. 
Further work into the prospects of binary asteroid survival throughout stellar evolution in the lead up to the white dwarf phase could help further elucidate the role that asteroid binarity has in post-main-sequence planetary systems.

\section*{Acknowledgements}
We thank the anonymous reviewer for their careful and insightful comments which have greatly improved this manuscript.
Simulations in this paper made use of the \texttt{REBOUND} N-body code \citep{ReinLiu2012}. The simulations were integrated using WHFast, a symplectic Wisdom-Holman integrator \citep{ReinTamayo2015, WisdomHolman1991}. The SimulationArchive format was used to store fully reproducible simulation data \citep{ReinTamayo2017}. 

DV gratefully acknowledges the support of the STFC via an Ernest Rutherford Fellowship (grant ST/P003850/1).

\section*{Data Availability}

The code used to produce the simulations discussed in this work, and the SimulationArchive files are available upon reasonable request to the corresponding author.


\bibliographystyle{mnras}
\bibliography{references} 




\bsp	
\label{lastpage}
\end{document}